\documentclass[twocolumn]{aastex701}

\begin{document}

\title{``Helium Enrichment Law": The Relation Between Helium and Metal Enrichment in Stars in FIRE-3 Simulations of Milky Way-mass Galaxies}

\shorttitle{``Helium enrichment law" in simulated MW-mass galaxies}
\shortauthors{Pratik J. Gandhi et al.}

\correspondingauthor{Pratik J. Gandhi}
\
\email{pratik.gandhi@yale.edu, gandhipratik1995@gmail.com}

\author[orcid=0000-0003-0965-605X, gname='Pratik', sname='Gandhi']{Pratik J. Gandhi}
\affiliation{Department of Astronomy, Yale University, New Haven, CT 06520, USA}
\email{pratik.gandhi@yale.edu}  

\author[orcid=0000-0003-4456-4863, gname='Earl', sname='Bellinger']{Earl P. Bellinger}
\affiliation{Department of Astronomy, Yale University, New Haven, CT 06520, USA}
\email{earl.bellinger@yale.edu}

\author[orcid=0000-0003-0603-8942, gname='Andrew', sname='Wetzel']{Andrew Wetzel}
\affiliation{Department of Physics \& Astronomy, University of California, Davis, CA 95616, USA}
\email{awetzel@ucdavis.edu}

\author[orcid=0000-0001-6380-010X, gname='Jeremy', sname='Bailin']{Jeremy Bailin}
\affiliation{Department of Physics \& Astronomy, University of Alabama, Tuscaloosa, AL 35487-0324, USA}
\email{jbailin@ua.edu}

\author[orcid=0000-0002-6163-3472, gname='Sarbani', sname='Basu']{Sarbani Basu}
\affiliation{Department of Astronomy, Yale University, New Haven, CT 06520, USA}
\email{sarbani.basu@yale.edu}

\author[orcid=0000-0002-4872-1021, gname='Isabella', sname='Trierweiler']{Isabella L. Trierweiler}
\affiliation{Department of Astronomy, Yale University, New Haven, CT 06520, USA}
\email{isabella.trierweiler@yale.edu}

\author[orcid=0000-0002-5908-737X, gname='Francisco', sname='Mercado']{Francisco J. Mercado}
\affiliation{Physics \& Astronomy Department, University of Southern California, Los Angeles, CA 90007, USA}
\email{fm94826@usc.edu}



\begin{abstract}

Helium abundance is a key input to stellar models that are used to infer accurate ages, masses, and compositions for individually resolved stars in the Milky Way (MW), underpinning Galactic archaeology. Since spectroscopically measuring He abundances is difficult, stellar models often assume a fixed ``chemical enrichment law" $\Delta Y$/$\Delta Z$ (the slope linearly relating a star's helium mass fraction $Y$ to its metal mass fraction $Z$). If incorrect, this assumption can introduce significant systematics in inferring stellar properties. Using the FIRE-3 cosmological baryonic zoom-in simulations, we investigate helium versus metal enrichment in the stars of MW-mass galaxies. Given the nucleosynthetic yields in FIRE-3, overall metal production is dominated by core-collapse supernovae (CCSN) at all progenitor metallicities. He production is dominated by CCSN for progenitor metallicities of $Z \lesssim 0.1 Z_\odot$, but by stellar winds for progenitors near $Z_\odot$. Our simulations produce a range of $\Delta Y$/$\Delta Z$ slopes slopes with best-fit values of $\approx 0.75 - 2.1$ depending on the age of the stars considered at present day; on average consistent with the range of values from MW observations but with no single ``universal" enrichment law. The $\Delta Y$/$\Delta Z$ slope also shows galaxy-to-galaxy variance in our simulations, and to leading order increases over cosmic time within the same galaxy, based on the peculiarities of the formation histories of each MW-mass system. Despite choosing a fixed set of nucleosynthetic yields for stellar feedback processes in FIRE-3, we predict variance in $\Delta Y$/$\Delta Z$ for stars in MW-mass galaxies, and observations of MW stars must be interpreted in this context.

\end{abstract}



\section{Introduction}
\label{sec:1-intro}

After the Big Bang and before the formation of the first stars, the elemental composition of the Universe was primarily hydrogen (H) and helium (He), with very trace amounts of Lithium (Li) and Beryllium (Be) as per Big Bang Nucleosynthesis (BBN) models \citep[e.g.,][]{steigman-10}. Heavier elements were produced via stellar processes, and the cosmos was enriched in metals over time as the result of stellar feedback; see \citet{kobayashi-20} for a comprehensive review on the origins of various heavy elements. The elemental abundances of stars thus encode a wealth of information about the formation and evolution of their host galaxies, and the conditions in the interstellar medium (ISM) at various points in cosmic time. The modelling of the elemental evolution of galaxies and in particular, the Milky Way (MW), is the subject of a rich field; see for example \citet{tinsley-80}, \citet{matteucci-21}, and \citet{somerville-15}.

The MW is the best studied galaxy \citep[see][for an excellent review]{bland-hawthorn-16} thanks to our ability to individually measure the properties of millions-to-billions of its constituent stars at present day. A key goal of ``Galactic Archaeology" of the MW is is to use the observed dynamics, ages, and elemental abundances of its stars at present day to uncover the Galaxy's formation history, accretion or assembly history, and shed light on the astrophysical processes governing galaxy formation \citep[as in][]{feast-00, freeman-02, sandage-06, belokurov-13, rix-13, catchpole-16, helmi-20, rix-22, deason-24, imig-25}. Galactic archaeology has been revolutionized by recent large-scale astrometric, spectroscopic, and asteroseismic surveys like \textit{Gaia} \citep{gaia-16}, APOGEE \citep{majewski-17}, LAMOST \citep{cui-12}, GALAH \citep{de-silva-15}, SDSS-V Milky Way Mapper \citep{sdss-26}, \textit{Kepler} \citep{borucki-10}, and TESS \citep{ricker-15}; in particular via measurements of the ages, elemental abundances, and distances of red giant branch (RGB) stars that are bright enough to be observed across the entire Galaxy.

Detailed stellar evolution models underpin the interpretation of these resolved observations of RGB stars in the MW. The primary inputs to such stellar models are the star's mass, [Fe/H] metallicity, and initial helium abundance \citep[for example,][]{bressan-12, dotter-16, spada-17, hidalgo-18, dotter-26}. Although we can spectroscopically measure [Fe/H] for RGB stars, we can only measure spectroscopic He abundances for hot stars ($\gtrsim 12000$ K) and not RGB, because cooler stars do not reach the temperatures necessary to produce optical He transitions and their spectra have weak (or nonexistent) He lines \citep[see for example][]{tayler-67}. 

Some studies have inferred He abundances in star clusters or stellar populations using isochrone fitting \citep[for example,][]{casagrande-07}. However, for Galactic archaeology we often want to know the ages of field stars, requiring their He abundances as inputs to stellar models. Asteroseismology offers a way to measure the ages of field stars, but even state-of-the-art seismic models still depend on an assumed linear scaling between He and heavy element/metal abundances, anchored to Big BBN and Solar calibrations \citep[see reviews by][]{chaplin-13, hekker-17}. The standard approach is to apply a transformation from the observed metallicity [Fe/H] to fractional abundances of heavy elements $Z$ based on Solar abundance ratios, and then on to fractional helium abundance $Y$ by assuming a relation between metal and helium content known as the ``chemical enrichment law" $\Delta Y$/$\Delta Z$, anchored to $Y$ and $Z$ values obtained from BBN \citep{steigman-10, planck-20}. Thus if the assumption of a fixed scaling between He and metal content for all stars in the MW is inaccurate, it can introduce systematics when inferring stellar properties (especially age) using standard stellar evolution or asteroseismic models. The existence and value of such a ``chemical enrichment law" have long been debated, and current observations favor values in the range $0.8 \lesssim \Delta Y$/$\Delta Z \lesssim 3$ as obtained from different measurements of individual MW stars \citep[][and references therein]{hacyan-76, chiosi-82, lebreton-99, jimenez-03, balser-06, casagrande-07, serenelli-10, silva-aguirre-17, verma-19}. Since accurately measuring stellar ages is key for Galactic archaeology, assuming such an enrichment law $\Delta Y$/$\Delta Z$ \textit{can significantly impact the inferences for MW science} when applying stellar models to asteroseismic observations of RGB stars.

It is important to note that the assumption of a fixed slope $\Delta Y$/$\Delta Z$ implies that over the course of the a galaxy's formation history, helium is created (or destroyed) in lock-step with metals. Given that helium has at least two primary modes of production --- core-collapse supernovae (CCSN) and stellar winds from asymptotic giant branch (AGB) stars --- with different yields \citep[][]{kobayashi-26}, there is no particular \textit{a priori} reason that this should hold. Additionally, many stars in the MW \textit{do not} have scaled version of Solar abundances \citep[][]{bland-hawthorn-16, imig-23}, which is another reason to not expect all elements to scale in lock-step over the course of the Galaxy's formation history.

Recent asteroseismic efforts have attempted to independently infer the He abundances of field stars using seismic glitch analyses and evolutionary models \citep[][]{gough-90, basu-04, broomhall-14, verma-14, corsaro-15, silva-aguirre-17, verma-19, saunders-23}. These results suggest a wide range of $\Delta Y$/$\Delta Z$ values for MW stars, potentially indicating that there is no single enrichment law for He that applies across the entire Galaxy. This motivates the use of theoretical models of galaxy formation and evolution with self-consistent treatment of star formation, stellar feedback, and elemental enrichment, to understand the evolution of stellar He relative to metals over the MW's formation history.


Specifically, there are very few studies so far \citep[to our knowledge, only][]{vincenzo-19} that leverage cosmological simulations of galaxy formation to study helium enrichment relative to heavy elements in galaxies like the MW. Advancements in numerical techniques, computing power, and astrophysical modelling over the last $\approx 15$ years have made possible cosmological zoom-in simulations that model the co-evolution of dark matter (DM), gas, and stars as well as star formation, stellar feedback, and elemental enrichment, all at high resolution. Multiple simulation projects existing today can form MW-mass galaxies with well-ordered, rotating disks and realistic star formation histories (SFHs) and elemental abundances; including Auriga \citep{grand-17, grand-24, iza-24}, ChaNGa \citep{applebaum-21}, NIHAO \citep{buck-20, buck-23}, Vintergatan \citep{agertz-21, renaud-21, renaud-25}, and FIRE \citep{hopkins-18b, hopkins-23, wetzel-23, wetzel-25}. 

In this work, we leverage cosmological baryonic zoom-in simulations from the FIRE project to analyze the evolution and enrichment patterns of stellar helium and heavy element abundances in MW-mass galaxies. We investigate whether the simulations support the idea of a single galaxy-wide ``chemical enrichment law", to test whether this is a justified assumption for stellar evolution and asteroseismic models. Throughout, we refer to the stellar helium mass fraction ($Y$) and heavy element or metal mass fraction ($Z$), defined as the ratio of helium mass to total mass and the ratio of mass of all elements heavier than helium to the total mass, respectively. Along with the hydrogen mass fraction ($X$), we require that the following condition is always satisfied, by definition:
\[
X + Y + Z \equiv \frac{\rm{m_H}}{\rm{m_{total}}} + \frac{\rm{m_{He}}}{\rm{m_{total}}} + \frac{\rm{m_Z}}{\rm{m_{total}}} = 1.
\]
\vspace{0.5mm}

\section{Methods} 
\label{sec:2-methods}

\subsection{Overview of FIRE-3 simulations}
\label{sec:2.x-fire}

We generated and analyzed cosmological baryonic zoom-in simulations from the Feedback in Realistic Environments (FIRE) project\footnote{\url{https://fire.northwestern.edu/}}, run with the massively parallel multi-methods \textsc{Gizmo} code\footnote{\url{http://www.tapir.caltech.edu/~phopkins/Site/GIZMO.html}} in its meshless finite-mass (MFM) mode \citep{hopkins-15}. 
In particular, our sample comprises three high-resolution zoom-in simulations --- \texttt{m12i, m12q, m12b} --- centered on cosmologically isolated MW-mass galaxies run with FIRE-3 physics, with halo masses of $M_{200 \rm m} = 1-2 \times 10^{12}\,\rm{M_\odot}$ and stellar masses comparable to the MW \citep{bland-hawthorn-16}. All simulations in this work have a high DM mass resolution of $ 3.5\times10^5\,\rm{M_\odot}$, and a baryonic (gas/star) particle mass resolution  $ 7070\,\rm{M_\odot}$. The gravitational force softening (proxy for spatial resolution) for DM and star particles is fixed at $\epsilon_{\rm DM} = 40$ pc and $\epsilon_{\rm star} = 4$ pc. The force softening and smoothing length for gas cells are adaptive down to a minimum of $1$ pc. At average inter-stellar medium (ISM) densities of $\approx 1 \, \rm{cm}^{-3}$, the gas softening is $\approx 40$ pc.  

The FIRE simulations employ state-of-the-art numerical methods and physics prescriptions for fluid dynamics, gas heating and cooling, star formation, and stellar feedback. The simulations explicitly resolve the dense, clumpy, multi-phase nature of the ISM \citep{hopkins-18b, hopkins-23} as opposed to applying an effective equation-of-state model. For full details on numerical and physical prescriptions of stellar feedback in FIRE, refer to \citet{hopkins-18b, hopkins-18a, hopkins-23, hopkins-25}. FIRE also includes radiative gas heating and cooling across $1 - 10^{10}$ K as well as tracking $11$ independently-evolved elements: H, He, C, N, O, Ne, Mg, Si, S, Ca, Fe. The multi-channel stellar feedback physics in FIRE includes detailed models for: (i) mechanical feedback including continuous stellar mass loss from O/B and AGB stellar winds, core-collapse (Type II) and white dwarf (Type Ia) supernovae, and (ii) non-mechanical feedback including photoionization and photoelectric heating, and radiation pressure. 

We use the latest version of the FIRE-3 physics model, featuring key updates to star formation, stellar feedback, and gas physics \citep{hopkins-23} compared to its previous iteration, FIRE-2. The model uses an updated spatially uniform, meta-galactic UV/X-ray background \citep{faucher-giguere-20} to account for the global effects of cosmic reionization. Note that we \textit{do not} model magnetic fields/magneto-hydrodynamics or explicit cosmic ray transport in any of our simulations discussed in this paper, but we do not expect the lack of either of those to have significant impact on the evolution of galaxy-wide helium and heavy metal abundances that form the focus of our analysis \citep[see Figure 8 of][]{hopkins-23}.

Star particles in FIRE-3 are formed from parent gas particles once they meet all the criteria for star formation: Jeans unstable, self-gravitating, and in a converging flow. Unlike the previous criteria in FIRE-2, there is no arbitrarily density threshold (previously $n \gtrsim 1000 \, \rm{cm}^{-3}$) or molecular requirement for star forming gas. The resultant star particle has the same mass as the parent gas particle and inherits all its elemental abundances directly. \textit{FIRE-3 (like FIRE-2) models every star particle as a simple stellar population (SSP) with a single age and element abundance composition}, assuming a \citet{kroupa-01} initial mass function (IMF) for stars in a mass range of $0.1-100 \, \rm{M_\odot}$. For each star particle, we account for stellar evolution using isochrones from the January 2021 version of STARBURST99 \citep{leitherer-14}. Our input models for continuous mass loss from O/B + AGB stellar winds as well as core-collapse supernova (CCSN) rates are also from STARBURST99. For white dwarf or Type Ia supernova (WDSN) rates, we use the constraints from \citet{maoz-17}. At each time step for a star particle, we use the tabulated supernova rate models to determine the probability of an individual CCSN or WDSN occurring, and we explicitly time-resolve individual supernova explosions rather than considering their collective effects over a period of time. For continuous stellar feedback channels (stellar winds, photoionization and photoelectric heating, and radiation pressure), we consider the amount of mass, metals, energy, and momentum ejected at each time step for a star particle based on the assumed stellar evolution models.    

Finally, on the numerical side, FIRE-3 leverages updates to the tree-based gravity solver with more accurate opening and time-stepping criteria \citep{grudic-20, hopkins-23b}.

The zoom-in initial conditions \citep{onorbe-14} for the simulations we analyze are derived from periodic cosmological parent boxes with side lengths of $86-172$ Mpc, and were generated at $z=100$ using the MUSIC code \citep{hahn-11}. All our simulations assume a flat $\Lambda$CDM cosmology, with parameters broadly consistent with \citet{planck-20}: $h = 0.68 - 0.71$, $\Omega_{\Lambda} = 0.69 - 0.734$, $\Omega_{\rm m} = 0.266-0.31$, $\Omega_{\rm b} = 0.0455-0.048$, $\sigma_8 = 0.801-0.82$, and $n_s = 0.961 - 0.97$. Table~\ref{tab:sims} lists the FIRE-3 zoom-in simulations analyzed in this work along with the basic properties of the MW-mass galaxy in each one. 

\begin{table*}
\centering
\caption{
Properties of the FIRE-3 MW-mass simulations in our sample.}
\vspace{-2 mm}
\addtolength{\tabcolsep}{7 pt}
\begin{tabular}{ccccccc}
\hline
\hline
\vspace{1mm}
Simulation & Baryonic mass & $M_{\rm star}$ of MW & $M_{200 \rm m}$ of MW & $R_{200 \rm m}$ of MW & Ref.$^{\dagger}$\\
name & resolution & at $z = 0$ & at $z = 0$ & at $z = 0$ & \\
& [$\rm{M}_\odot$] & [$\times 10^{10} \; \rm{M}_\odot$] & [$\times 10^{12} \; \rm{M}_\odot$] & [kpc] & \\
\hline
\texttt{m12q} & $7070$ & $5.6$ & $1.6$ & $371$ & \citet{steel-26}\\
\texttt{m12i} & $7070$ & $3.4$ & $1.7$ & $380$ & this work\\
\texttt{m12b} & $7070$ & $2.1$ & $1.4$ & $358$ & this work\\
\hline
\hline
\end{tabular}
\addtolength{\tabcolsep}{-7 pt}
\tablecomments{
Our FIRE-3 simulations in descending order of stellar mass of the central MW-mass galaxy. ``Baryonic mass resolution" lists the masses that gas particles are initialized with at the beginning of the simulation. We measure $M_{\rm star}$ for each galaxy within a spherical volume of radius $15$ kpc. As a proxy for virial mass we use $M_{\rm 200m}$, the total mass within $R_{\rm 200m}$ defined as the radius out from the center of the DM halo containing $200$ times the mean matter density of the Universe.
The final column lists the paper that first introduced the simulation at this resolution with FIRE-3 physics (but no MHD).
}
\label{tab:sims}
\end{table*}

\subsection{Tracking elemental abundances in the simulations, including Y and Z}
\label{sec:2.x-measuring-elements}

The FIRE-2 simulations relied on the older \citet{anders-89} reference values for Solar elemental abundances, with $Z_\odot \approx 0.02$. In FIRE-3 and this work, we scale to the newer \citet{asplund-09} proto-Solar compositions, with mass fractions of \{$Z$, He, C, N, O, Ne, Mg, Si, S, Ca, Fe\} = \{$0.0142$, $0.2703$, $0.00253$, $0.000741$, $0.00613$, $0.00134$, $0.000757$, $0.000712$, $0.000331$, $0.0000687$, $0.00138$\}. Additionally unlike in FIRE-2, in FIRE-3 we initialize all gas at purely primordial elemental abundances instead of an arbitrary metallicity floor. Thus gas cells are initialized with $X = 0.76$, $Y = 0.24$, and $Z = 0.00$ as per constraints on Big Bang Nucleosynthesis from \citet{planck-20} measurements. Given the $11$ elemental species tracked in FIRE, the heavy metal mass (or mass fraction $Z$) is by definition the sum of the masses (or mass fractions) of C, N, O, Ne, Mg, Si, S, Ca, and Fe. This is broadly a justified assumption since the total metal budget in stars on average is overwhelmingly dominated by these $11$ elements. For example, the measurements in \citet{asplund-09} attribute $\approx 99 \%$ of the entire metal mass fraction in the Sun to the sum of these elements.

Any time a star formation event occurs in the simulations, the resultant star particle inherits the elemental abundances of its (single) parent gas element. \textit{Importantly, the elemental abundances of star particles in FIRE-3 are locked in at the time of their formation and do not change throughout the rest of their lifetime,} as is the standard for cosmological zoom-in simulations of galaxy formation. However, this is not an issue for our analysis since we focus on the evolution of the \textit{initial} helium and metal abundances of stars formed over cosmic time, rather than their present-day $Y$ and $Z$ values. Thus, tracking nucleosynthetic enrichment of the ISM from stellar feedback is far more important for this than evolution in internal stellar abundances, which our simulations are well-designed to do.

The FIRE-3 simulations also include an explicit treatment of unresolved turbulent sub-grid diffusion of metals in gas \citep[introduced in][]{hopkins-17, su-17, escala-18}. This is necessary to account for the naturally occurring metal diffusion in the ISM due to turbulent eddies and fluid instabilities. Our model treats the diffusion and mixing of each element separately.

\subsection{Assumed nucleosynthetic yields in FIRE-3}
\label{sec:2.x-yields}

For a detailed discussion of nucleosynthetic yields in FIRE-3, refer to \citet{hopkins-23}; here we summarize the ideas most relevant to this work. We include important updates to the yields from continuous mass loss (stellar winds) in FIRE-3, namely the time dependence after the birth of a stellar population to account for the distinction between early-time O/B winds and late-time AGB winds, as well as incorporating dependence of wind yields on the metallicity of the parent star particle \citep{hopkins-23}. The yields we source from a combination of \citet{leitherer-14}, \citet{cristallo-15}, and \citet{limongi-18}, as sampled and interpolated in \citet{prantzos-18}. The distinction between O/B and AGB wind models is especially important for this paper since we expect early-time O/B winds to primarily source heavier elements (especially C, N, O) and late-time AGB winds to be the more significant contributor of He, as far as continuous mass loss is concerned.

FIRE-3 treats all instances of white dwarf or Type Ia supernovae (WDSN) as identical, with ejecta masses of $1.4\,\rm{M_\odot}$. For nucleosynthetic yields of WDSN, we use the updated synthesis of \citet{iwamoto-99} and \citet{leung-18}, as described in \citet{mori-18}. 

Although FIRE-2 treated all sampled core-collapse supernovae (CCSN) identically with IMF-averaged yields, we update this in FIRE-3 to account for the fact that CCSN at different times after the formation of a SSP correspond to different progenitor stellar masses \citep{hopkins-23}. Since each star particle represents a mono-age stellar population with a well-sampled Kroupa IMF, we model CCSN yields over time as a function of the age of the particle (that is, the age of the stellar population at the time the individual supernova event is sampled). For the CCSN yields, we rely on \citet{sukhbold-16} since they sample densely in progenitor stellar mass and span at least the entire range of CCSN progenitors our assumed IMF covers ($\sim 8-120\,\rm{M_\odot}$). Note that in FIRE-3 we \textit{do not} assume any progenitor-$Z$ dependence on the CCSN yields, because the various yield model compilations \citep[including][]{nomoto-13, pignatari-16, sukhbold-16, limongi-18} do not agree even on the qualitative sign of the $Z$-dependence (positive versus negative) or how strong the trends are for various element species.

\subsection{Time- and $Z$-dependence of helium and heavy element yields in FIRE-3}
\label{sec:2.x-yields-vs-time}

To contextualize the results of this paper, we must consider the cumulative nucleosynthetic yields from the three mechanical feedback channels (stellar winds and both kinds of supernovae) as a function of time after the birth of a star particle (i.e., simple stellar population) in the simulations. Although the supernova yields in FIRE-3 are not metallicity-dependent, the stellar wind yields are in terms of both their overall mass loss rate and the mass fractions of elements ejected. Thus the cumulative yields for an SSP end up being dependent on its metallicity ($Z$). 

\begin{figure*}
\centering
\begin{tabular}{c c c}
\includegraphics[width = 0.32\linewidth]{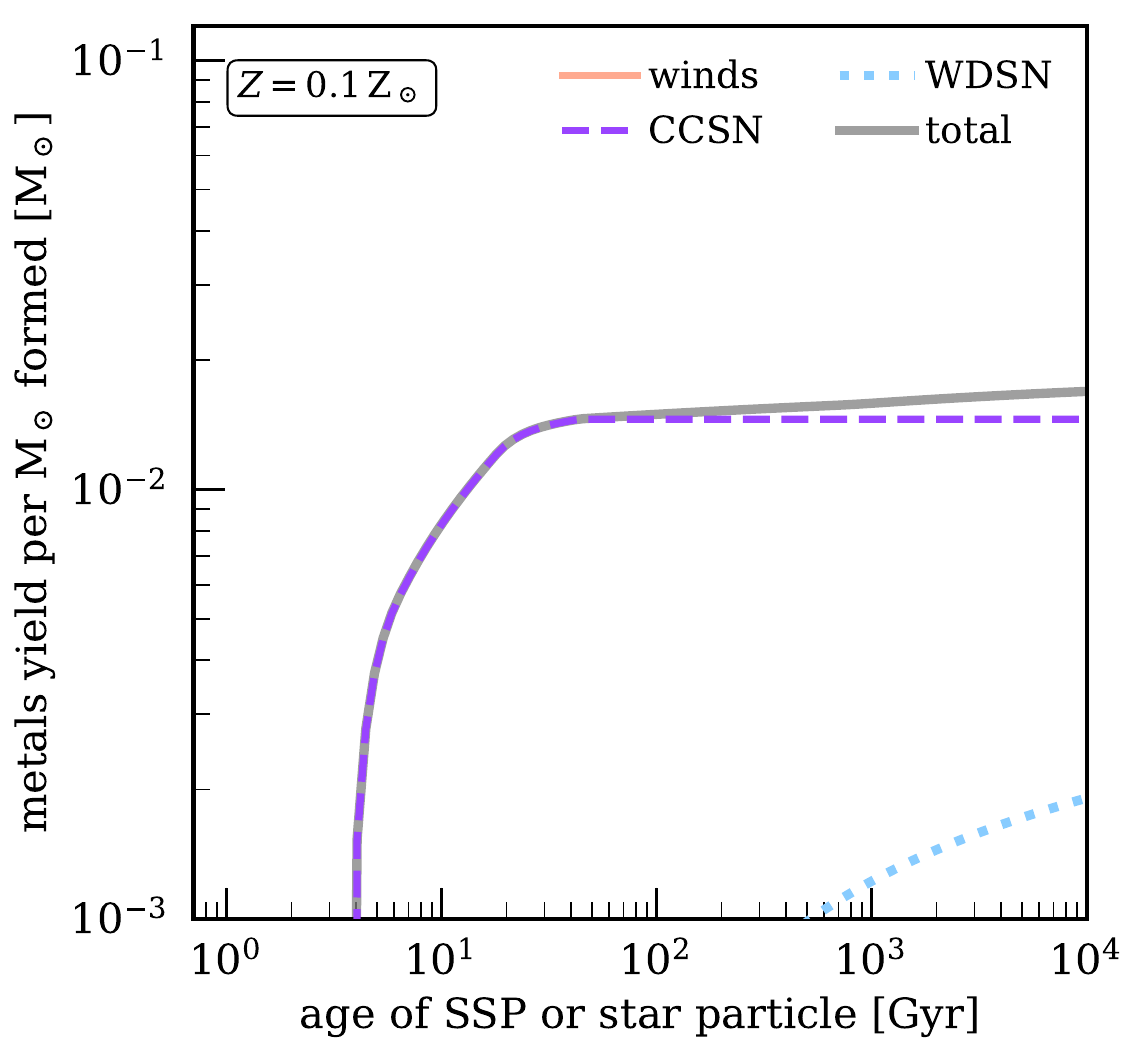} & 
\includegraphics[width = 0.32\linewidth]{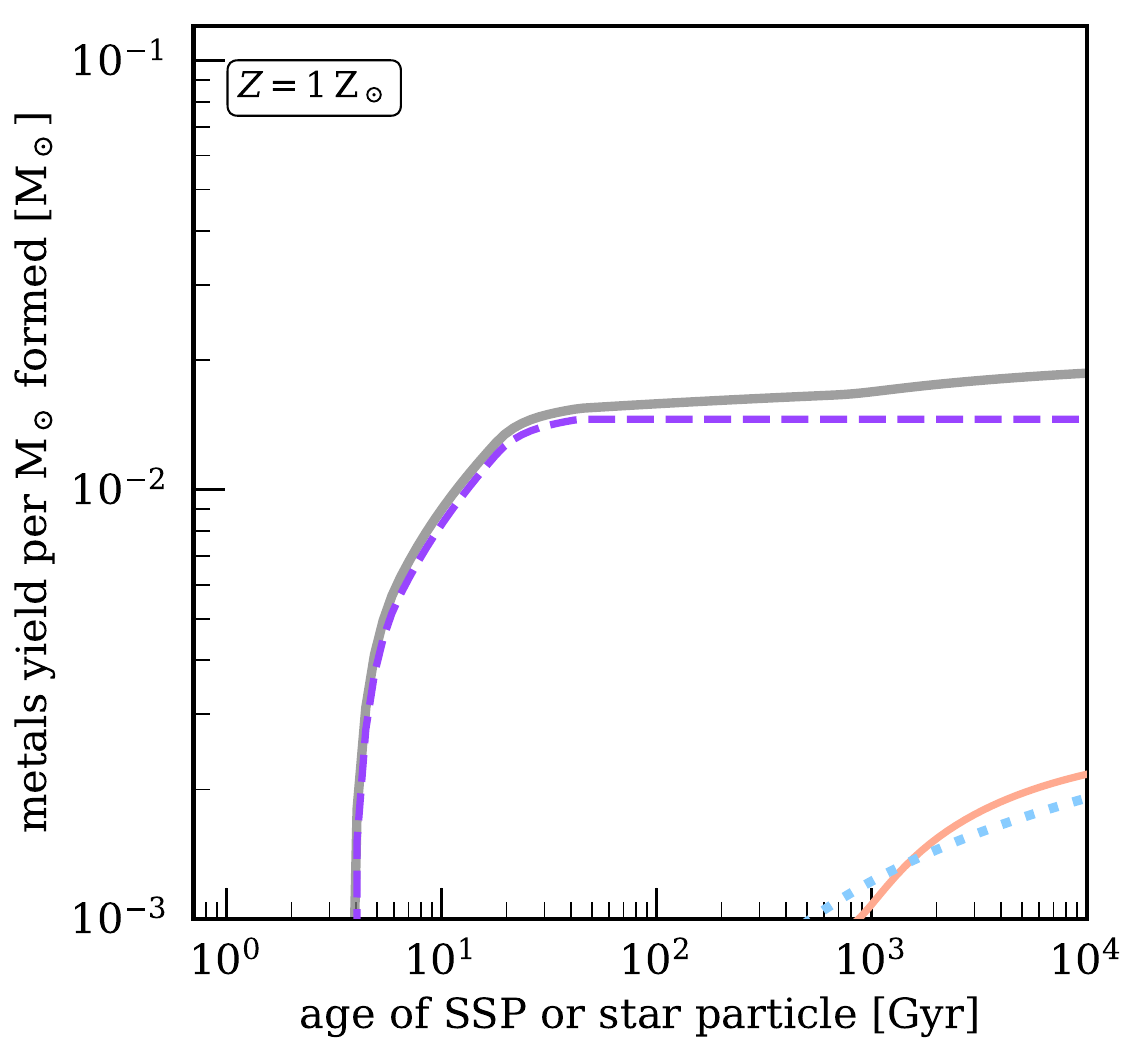} & 
\includegraphics[width = 0.32\linewidth]{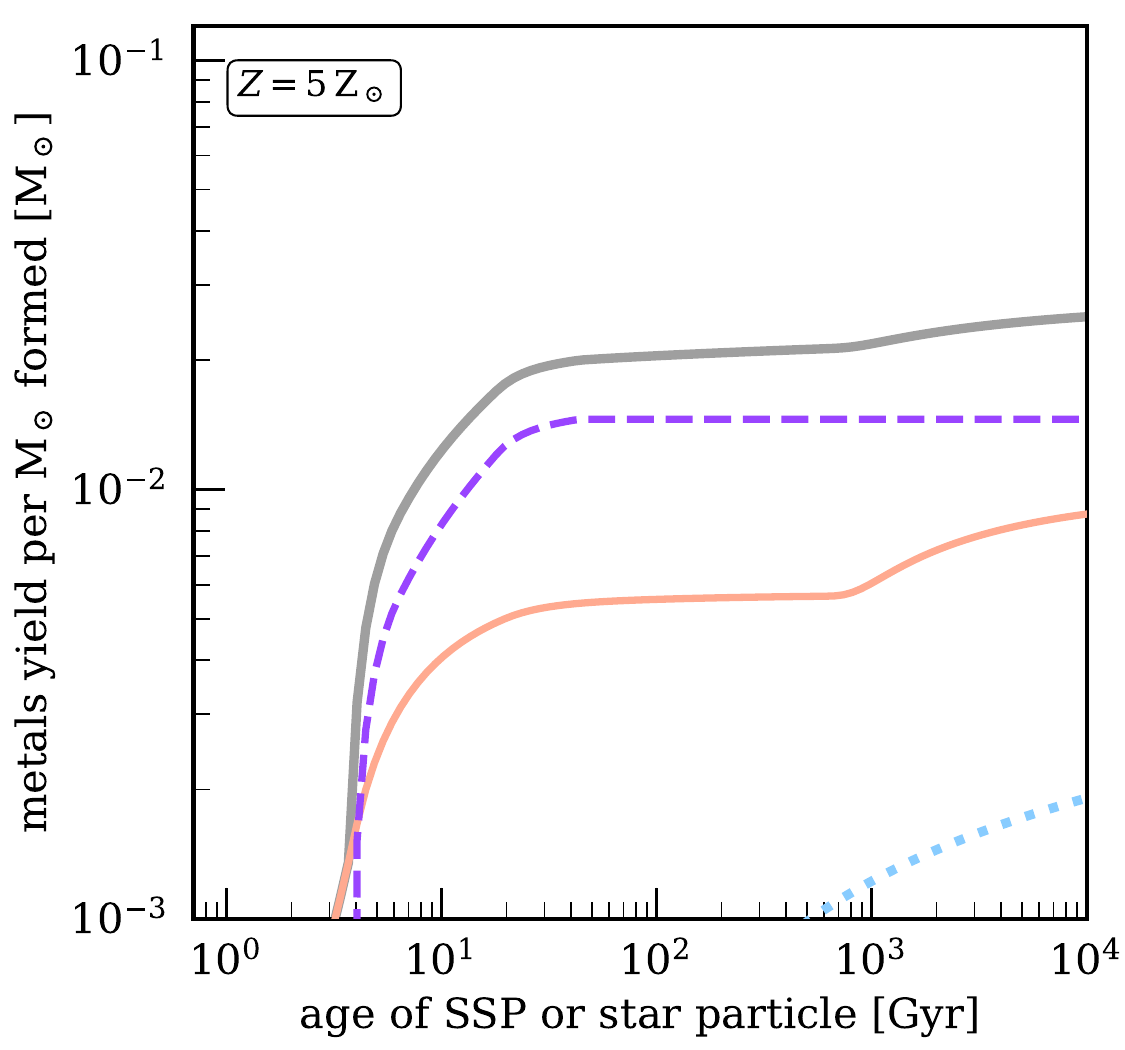} \\
\includegraphics[width = 0.32\linewidth]{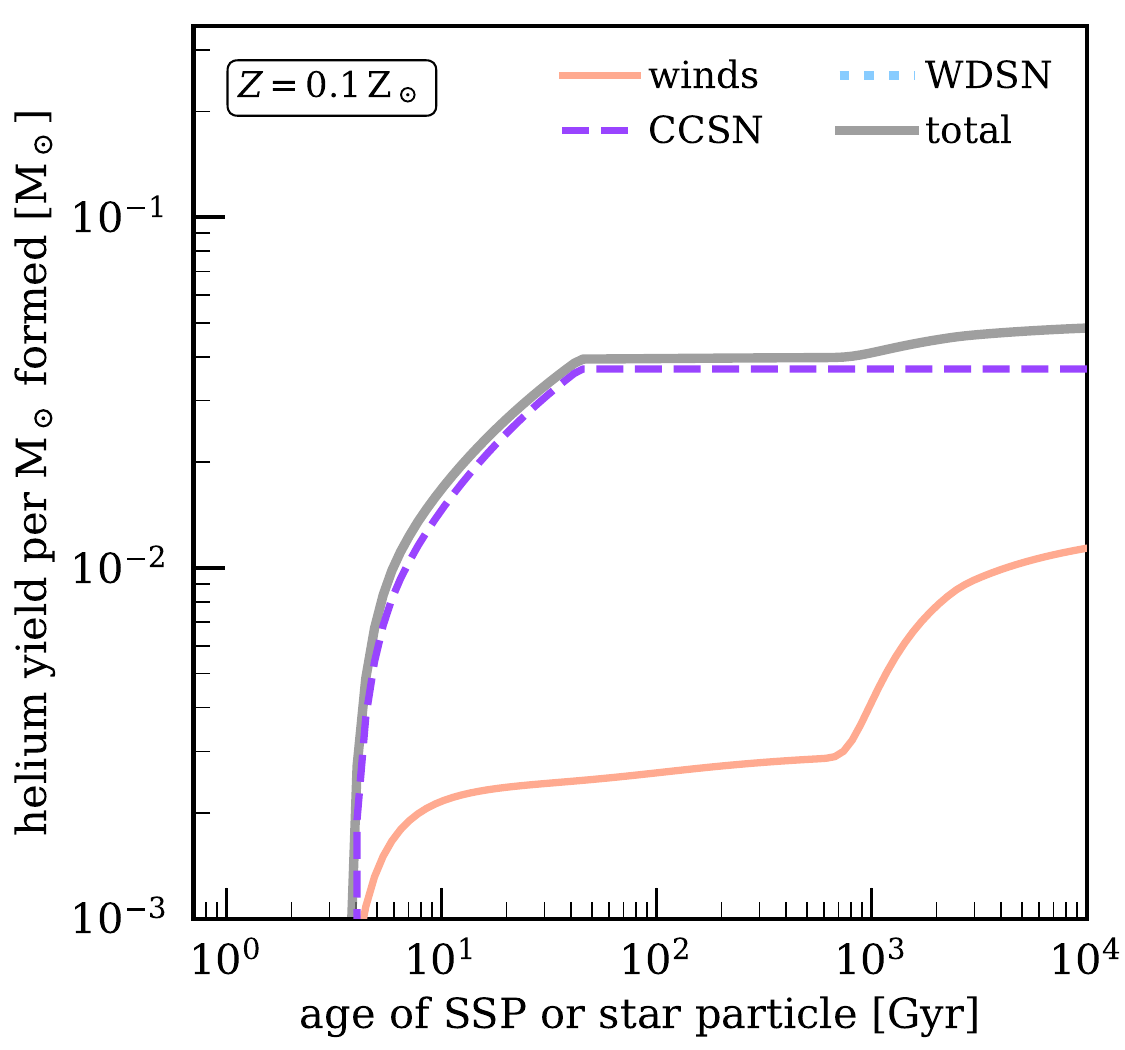} & 
\includegraphics[width = 0.32\linewidth]{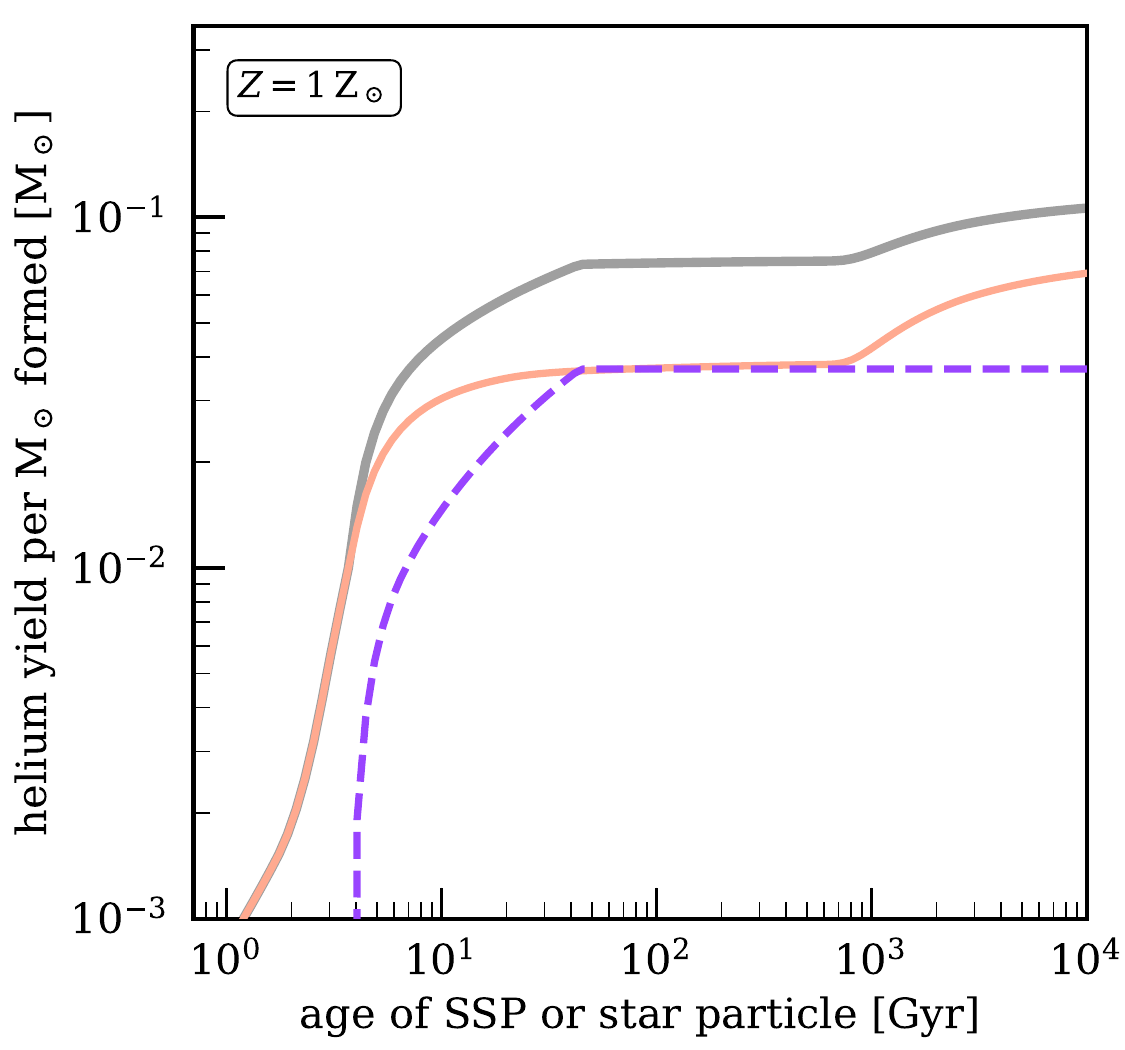} & 
\includegraphics[width = 0.32\linewidth]{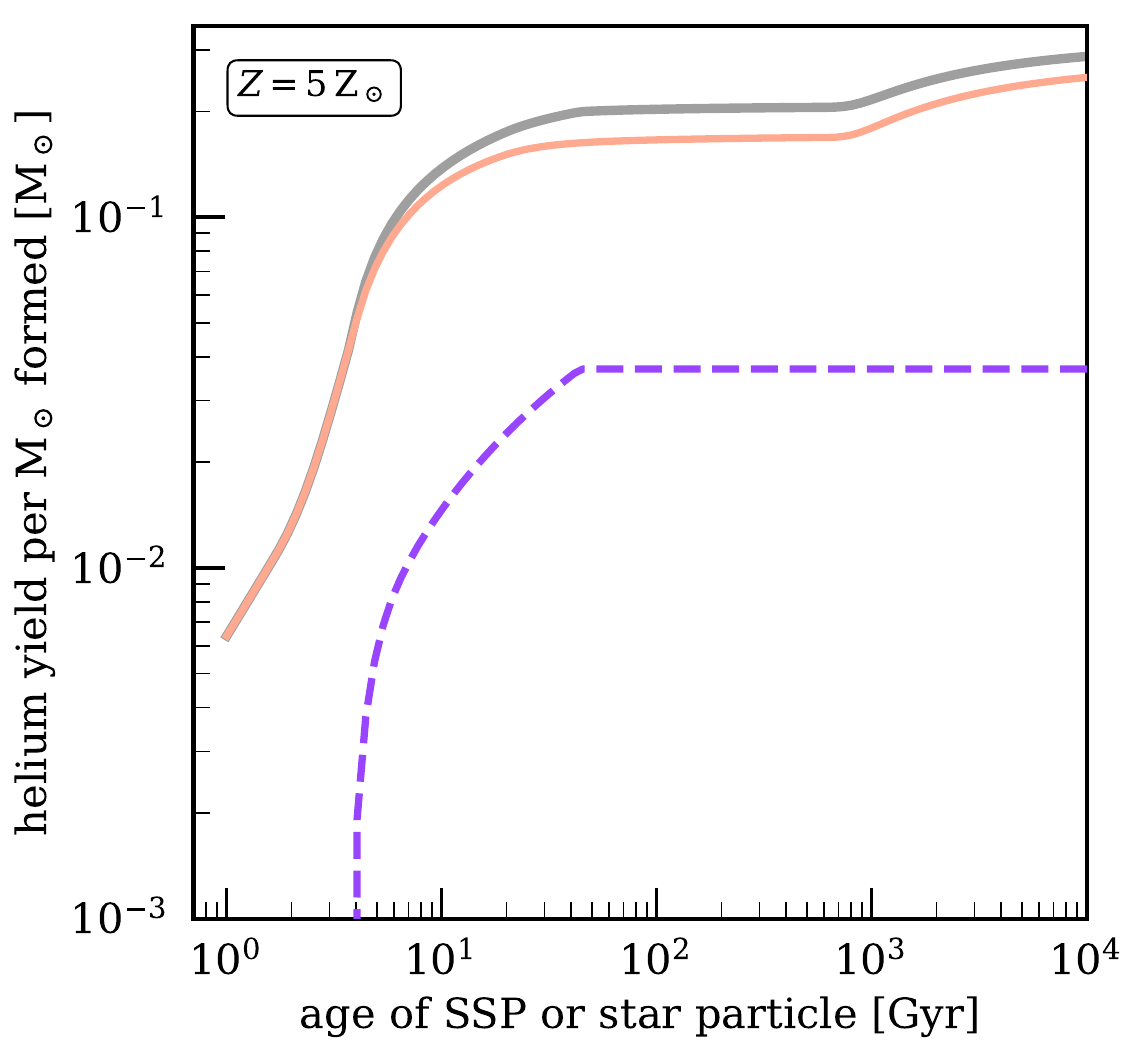}
\end{tabular}
\caption{
\textbf{Relative contributions of the three mechanical feedback channels in FIRE-3 to the helium and metal yields, as a function of time after star particle formation.} We show the yields from stellar winds, core-collapse supernovae, and white dwarf supernovae per unit $\rm{M}_\odot$ formed, as a function of age of the simple stellar population (SSP). Left panel shows results for an SSP at $0.1\rm{Z_\odot}$, middle for $1\rm{Z_\odot}$, and right for $5\rm{Z_\odot}$, to demonstrate how the relative contributions of the feedback channels evolve as a function of progenitor SSP metallicity. The top row shows how total heavy metal yields (for elements C through Fe) are dominated by core-collapse supernovae at all progenitor metallicities, and mostly saturate by $\approx250$ Myr after the birth of the SSP. The bottom row shows that core-collapse supernovae dominated helium yields at $\lesssim 0.1\,\rm{Z_\odot}$ and saturate within $\approx 400$ Myr, stellar wind helium yields become dominant at $\gtrsim 1\,\rm{Z_\odot}$, with an early-time contribution from O/B winds and a similar late-time (after $1$ Gyr) contribution from AGB winds. 
}
\label{fig:yields-vs-age}
\end{figure*}

Figure~\ref{fig:yields-vs-age} shows the contribution of each of the three feedback channels to the overall metal and helium yield per unit $\rm{M}_\odot$ of stars formed, as a function of time and varying based on progenitor metallicity. Recall here that the total metal yield represents the sum of C, N, O, Ne, Mg, Si, Ca, and Fe. The top row breaks down the contributions of stellar winds, CCSN, and WDSN to the overall metal yield versus time after birth of a stellar population, while the bottom row does the same for the helium yield. From left to right, the columns show how the yield contributions vary as a function of the metal fraction of the parent stellar population, at $Z = \{0.1, 1, 5\} \rm{Z_\odot}$. 

We show that for the FIRE-3 feedback prescriptions, the total metal yield is consistently dominated by CCSN; by at least an order of magnitude more than WDSN and winds for all values of $Z \leq Z_\odot$, and by $\approx 2\times$ for $Z = 5 Z_\odot$. Also, while stellar winds contribute the least to overall metal yields at lower $Z$, at near-solar values they contribute roughly the same as WDSN after a Hubble time ($\sim 10$ Gyr) and at super-solar values they contribute significantly more than WDSN after a Hubble time, with an early burst from O/B winds followed by a late-time increase from AGB winds after $\sim 1$ Gyr. Overall though, the total metal yield (everything except H and He) is dominated by CCSN throughout a galaxy's enrichment history, and this yield contribution is quite prompt after a burst of star formation, with about half of it within just $\sim 10$ Myr after the birth of an SSP.

Figure~\ref{fig:yields-vs-age} also shows that the He yield is dominated by prompt feedback from CCSN (within $\sim 400$ Myr) at $Z \leq 0.1\,\rm{Z_\odot}$ (by about half an order of magnitude compared to stellar winds after a Hubble time). However, for parent stellar populations at $Z\sim 1\,\rm{Z_\odot}$, the stellar wind He contribution exceeds that of CCSN at both early and late times, because of O/B winds and AGB winds respectively. The wind contribution to He yields at this parent stellar population metallicity ends up being roughly equal from early O/B winds and late AGB winds, after $\sim 10$ Gyr. Additionally, this stellar wind He contribution begins very soon after a burst of star formation --- within $\sim 2$ Myr --- owing to continuous mass loss from winds turning on right after star formation. Finally, for well over Solar parent populations ($Z\sim5Z_\odot$), the stellar wind contribution to the He yield is almost an order of magnitude higher than that of CCSN for all times after star formation. We also highlight that unlike the overall metal yield, WDSN contribute nothing to the He yield, which remains a balance between winds and CCSN. Since galaxies enrich in elements over cosmic time, stars forming earlier have much lower $Z$ and therefore different yield timings than those that form later. This has important implications for the timing of He versus metal enrichment, since the stellar wind contribution to He yields dominates over that of CCSN for parent SSPs at $\gtrsim Z_\odot$.

\begin{figure*}
\centering
\begin{tabular}{c c c}
\includegraphics[width = 0.315\linewidth]{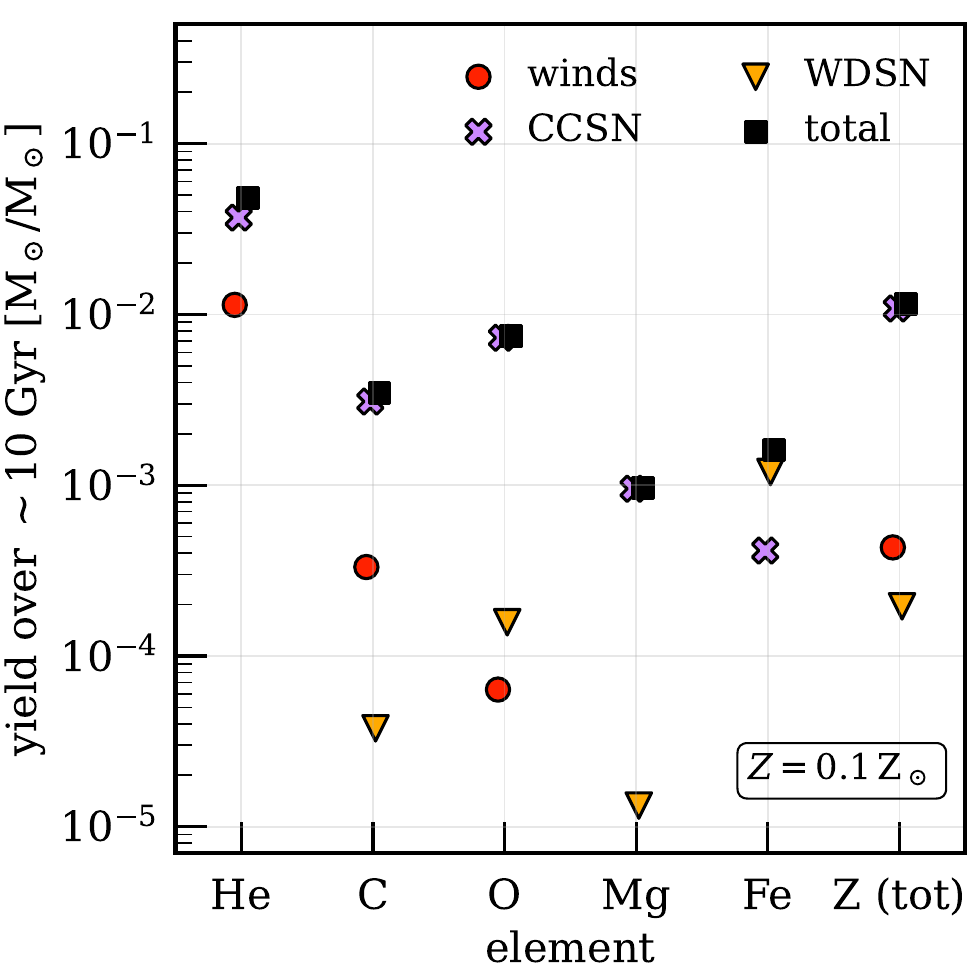} & 
\includegraphics[width = 0.315\linewidth]{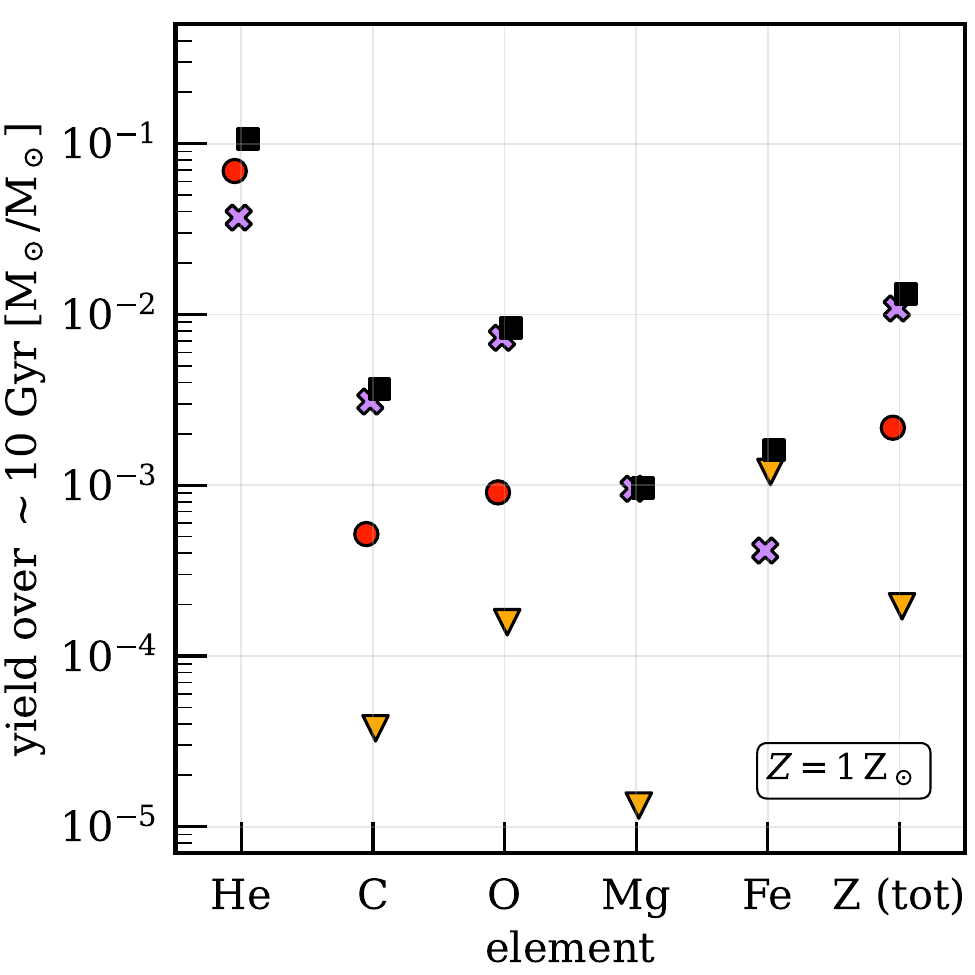} & 
\includegraphics[width = 0.315\linewidth]{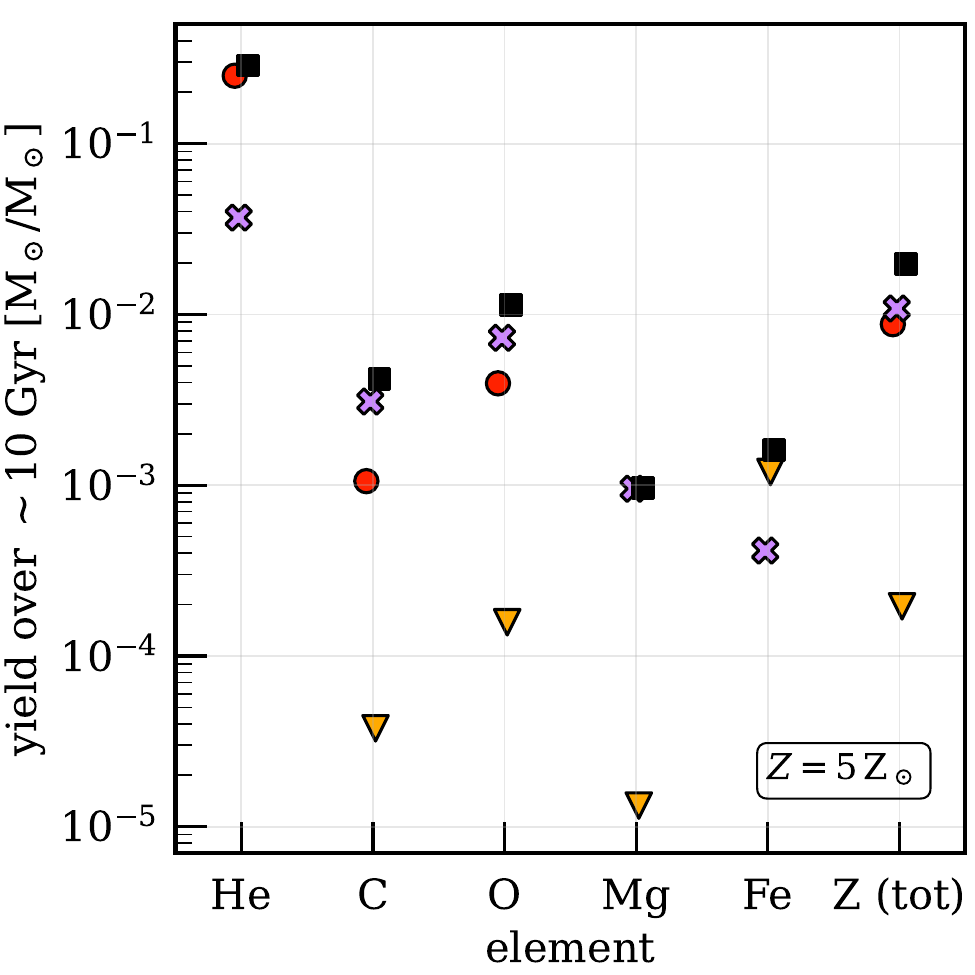}
\end{tabular}
\caption{
\textbf{Relative contributions of different FIRE-3 feedback channels to the cumulative nucleosynthetic yields over $\sim 10$ Gyr after a burst of star formation.} We show integrated yields for five select elements as well as total metals (for all $9$ heavy elements tracked by FIRE-3: C, N, O, Ne, Mg, Si, S, Ca, Fe) over $10$ Gyr for a simple stellar population (SSP). As in Figure~\ref{fig:yields-vs-age}, the left panel shows results for an SSP at $0.1\rm{Z_\odot}$, middle for $1\rm{Z_\odot}$, and right for $5\rm{Z_\odot}$. He production is dominated by core-collapse supernovae for $0.1 \rm{Z_\odot}$ but switches to being dominated by stellar winds (primarily late-time AGB winds, not early-time O/B winds) for $1-5 \rm{Z_\odot}$. The overall heavy metal yield (for elements C through Fe) is dominated by core-collapse supernovae at progenitor metallicities of $0.1 \rm{Z_\odot}$ and $1 \rm{Z_\odot}$, but switches to equal contributions from stellar winds and core-collapse supernovae at $5 \rm{Z_\odot}$.
}
\label{fig:channels-per-element}
\end{figure*}

Figure~\ref{fig:channels-per-element} shows the overall contribution of the three mechanical feedback channels (winds, CCSN, and WDSN) to the cumulative yields of various elements integrated over a Hubble time ($\sim 10$ Gyr) per unit $\rm{M_\odot}$ of stars formed. Again, from left to right each panel shows results for different metal mass fractions of the parent stellar population, at $Z = \{0.1, 1, 5\} \,\rm{Z_\odot}$. This figure highlights that at metallicities of $Z \leq 0.1\rm{Z_\odot}$, the cumulative He yields after $10$ Gyr are dominated by CCSN with roughly an order of magnitude lower contribution from stellar winds and no contribution from WDSN; at metallicities $\geq 1\rm{Z_\odot}$, the overall He yields from stellar winds is $\approx 2-8\times$ that from CCSN after about $10$ Gyr. The overall metal yield (for the $9$ elements ranging from C through Fe that are tracked in FIRE-3) is thoroughly dominated by CCSN at progenitor metallicities of $0.1\rm{Z_\odot}$ and $1\rm{Z_\odot}$ and switches to equal contributions from CCSN and winds at $5\rm{Z_\odot}$, despite WDSN being the biggest contributor to Fe in particular. Figure~\ref{fig:channels-per-element} demonstrates that it is important to self-consistently analyze the contributions of the different feedback channels and their relative timing when considering the enrichment of helium versus heavy metals in stars over a galaxy's formation history of a galaxy.

\subsection{Selection of stars in FIRE-3 galaxies for our analysis}
\label{sec:2.x-star-selection}

As our base sample, we consider at $z=0$ in each galaxy \textit{all stars} that lie within a sphere of radius $15$ kpc from the center of the galaxy. This selection effectively encompasses the thin disk, thick disk, bulge, and inner stellar halo of the simulated MW-mass galaxies at present day, useful for analyzing Galaxy-wide trends. 

Additionally, for our analysis of $Y-Z$ enrichment patterns and $\Delta Y$/$\Delta Z$ slopes, we have a \textit{secondary set of criteria} to approximate the selection of stars in the Solar neighborhood. This is useful because historically most observational analyses of helium and metal abundances in MW stars have focused on the Solar neighborhood \citep[e.g.,][]{casagrande-07, silva-aguirre-17, verma-19}, although that is poised to change with the \textit{Roman} and PLATO missions. We choose the following spatial cuts in cylindrical coordinates aligned with the disk plane of each galaxy, to select stars in the ``Solar annulus": $7.5\,\rm{kpc} \leq r \leq 8.5\,\rm{kpc}$ and $-1\,\rm{kpc} \leq z \leq 1\,\rm{kpc}$, where $r$ and $z$ represent the radial coordinate and vertical coordinate (off the mid-plane of the disk), respectively. At $z=0$, our simulated galaxies primarily show elemental abundance variations in the radial direction and not azimuthal variations across their disks, so our selection of a Solar annulus is a good approximation for the MW's Solar neighborhood, following the technique used in previous studies of FIRE simulations \citep{mccluskey-24, graf-25b, bellardini-26, steel-26}.

For both our samples --- (a) all stars within $15$ kpc and (b) only Solar annulus stars --- we select stars with metal mass fractions $Z \leq 0.05$ (that is, $\lesssim 3.5\,Z_\odot$), since there are no observed MW stars with higher abundances. In each simulated galaxy, ignoring stars with $Z > 0.05$ only removes $0.5\%$ of all stars, so this has a negligible effect on our results. Throughout our analysis, we do not distinguish between stars that formed \textit{in-situ} in the proto-MW and those that were accreted via mergers of lower-mass galaxies (or \textit{ex-situ}), primarily because observational analyses of $\Delta Y$/$\Delta Z$ in MW stars typically do not differentiate between in-situ and ex-situ populations. In future work, we aim to analyze helium and metal enrichment patterns in simulations of lower-mass galaxies, and their impact on the abundance patterns of ex-situ stars in the MW.

\section{Results} 
\label{sec:3-results}

\subsection{Distribution of stellar helium and heavy metal abundances}
\label{sec:3.x-Y-and-Z-PDFs}

\begin{figure*}
\centering
\begin{tabular}{c c}
\includegraphics[width = 0.45\linewidth]{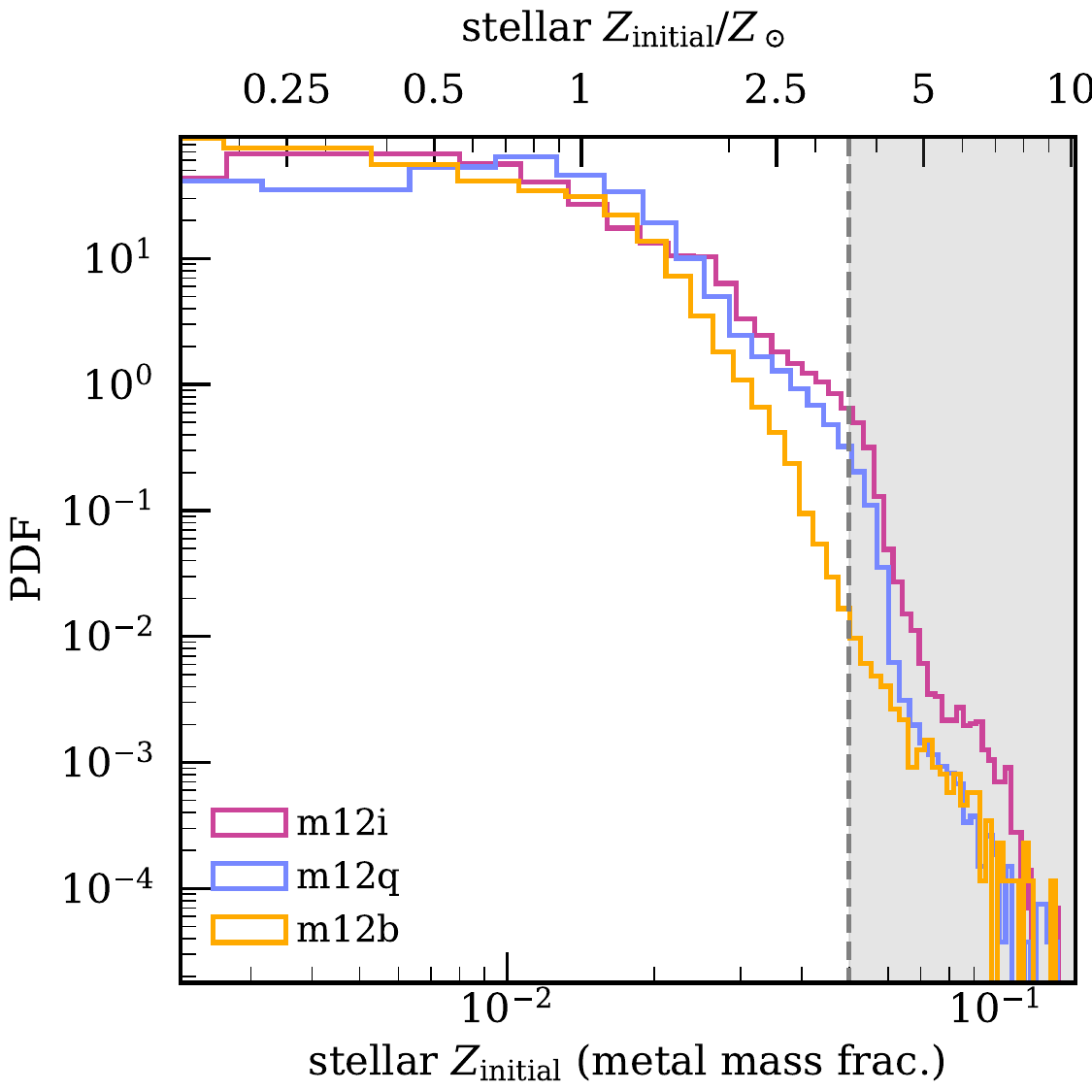} & \includegraphics[width = 0.45\linewidth]{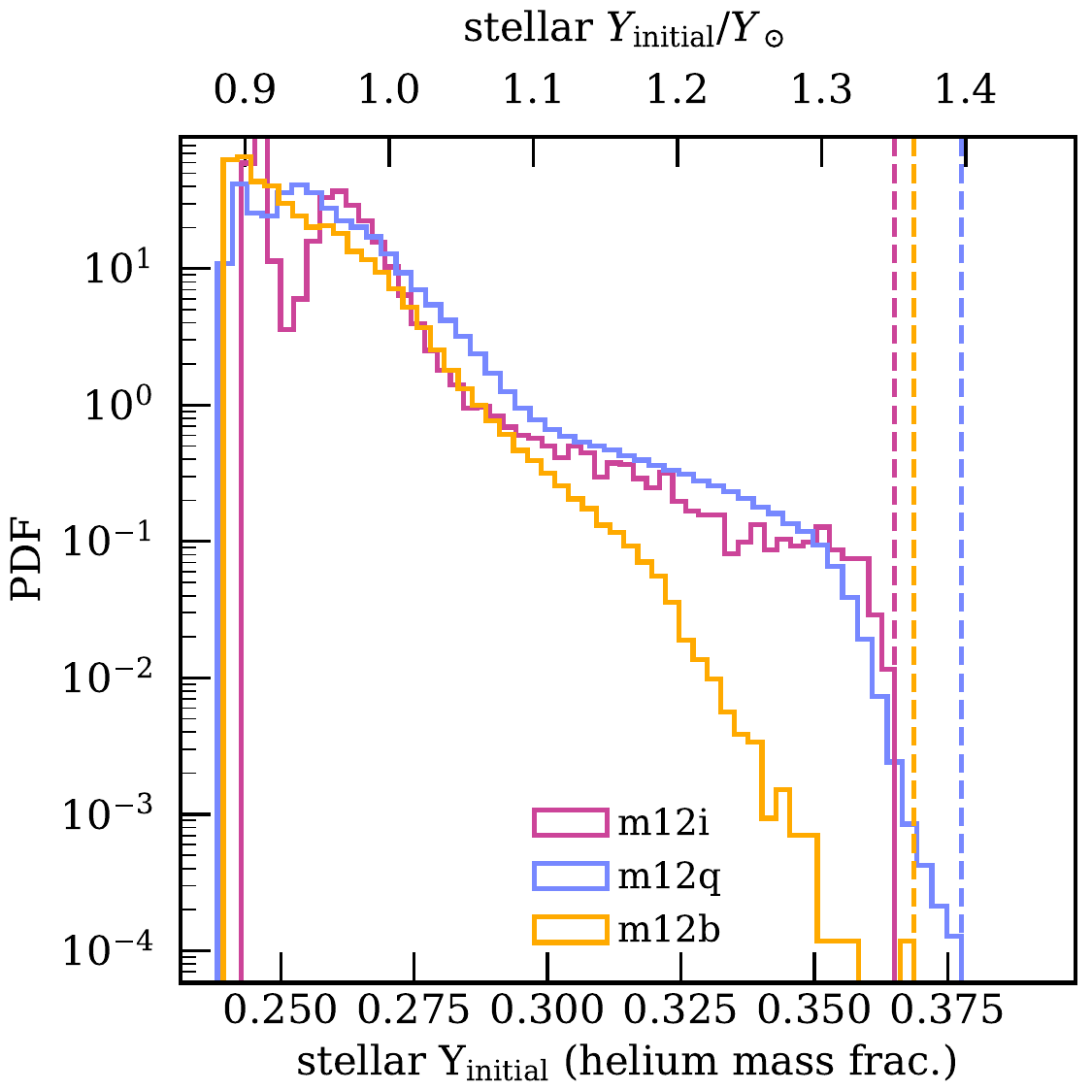}
\end{tabular}
\caption{
\textbf{Distribution of stellar $Z_{\rm initial}$ and $Y_{\rm initial}$ in our simulated galaxies at present day.} Left panel shows the PDF of initial $Z$ for all stars within a $15$ kpc sphere centered on each galaxy, and we ignore any stars with $Z_{\rm initial} > 0.05$ as shown by the grey shaded region (only $\sim 0.5\%$ of all stars). The PDFs for all three galaxies are consistent with each other to leading order, with similar median ($Z\approx0.006-0.011$, or, $\approx 0.4-0.8\,\rm{Z_\odot}$) values as described in Section~\ref{sec:3.x-Y-and-Z-PDFs}. On the right, we show the PDF of $Y_{\rm initial}$ values for all stars within $15$ kpc spheres with $Z_{\rm initial} \leq 0.05$, as identified in the left panel. Again, we find consistent distributions across the three simulated MW-mass galaxies, with similar median ($Y\approx0.248-0.255$, or, $\approx0.92-0.95\,\rm{Y_\odot}$) values. The vertical dashed lines show max($Y_{\rm initial}$): $0.365$ for \texttt{m12i}, $0.378$ for \texttt{m12q}, and $0.369$ for \texttt{m12b}.
}
\label{fig:Y-Z-PDFs}
\end{figure*}

We now consider, for the MW-mass galaxies in our sample, the distribution of stellar helium and metal abundances ($Y$ and $Z$) at present day. Figure~\ref{fig:Y-Z-PDFs} shows the probability distribution functions (PDFs) for the abundances of all stars in our galaxies within a sphere of radius $15$ kpc out from the center. To reiterate, star particles in FIRE-3 retain the elemental abundances of the gas element they form from, and then their abundances do not change over their lifetime despite accounting for stellar feedback. Hence, we show in Figure~\ref{fig:Y-Z-PDFs} the PDFs of $Z_{\rm initial}$ and $Y_{\rm initial}$ even though we are considering $z=0$. As the left panel shows, all three galaxies in our sample have similar distributions of stellar $Z_{\rm initial}$, with the median and $16-84$th percentile range values being $0.008$ for \texttt{m12i}, $0.011$ for \texttt{m12q}, and $0.006$ for \texttt{m12b}. The dashed vertical line highlights $Z = 0.05$ and the grey shaded region shows metal mass fraction values greater than that, which we \textit{do not} consider since there are no observed MW stars with $Z > 0.05$. 

In the right panel of Figure~\ref{fig:Y-Z-PDFs}, we show the PDFs for stellar $Y_{\rm initial}$, considering only the stars within a $15$ kpc sphere centered on each galaxy that have $Z \leq 0.05$ (based on the selection in the left panel). The three simulations again have broadly consistent distributions of helium abundances, with the median values being $0.249$ for \texttt{m12i}, $0.255$ for \texttt{m12q}, and $0.248$ for \texttt{m12b}. The dashed vertical lines show the maximum stellar $Y_{\rm initial}$ for each galaxy: $0.365$ for m12i, $0.378$ for m12q, and $0.369$ for m12b. These maximum $Y_{\rm initial}$ values are broadly consistent with the largest helium mass fractions measured in the most He-rich main sequence and red giant branch stars in the MW, especially those in globular clusters \citep[for example,][]{dantona-05, piotto-07}.

\subsection{Y versus Z abundance patterns for the present-day stars in FIRE-3 MW-mass galaxies}
\label{sec:3.x-Y-vs-Z-stars}

\begin{figure*}
\centering
\begin{tabular}{c c c}
\includegraphics[width = 0.32\linewidth]{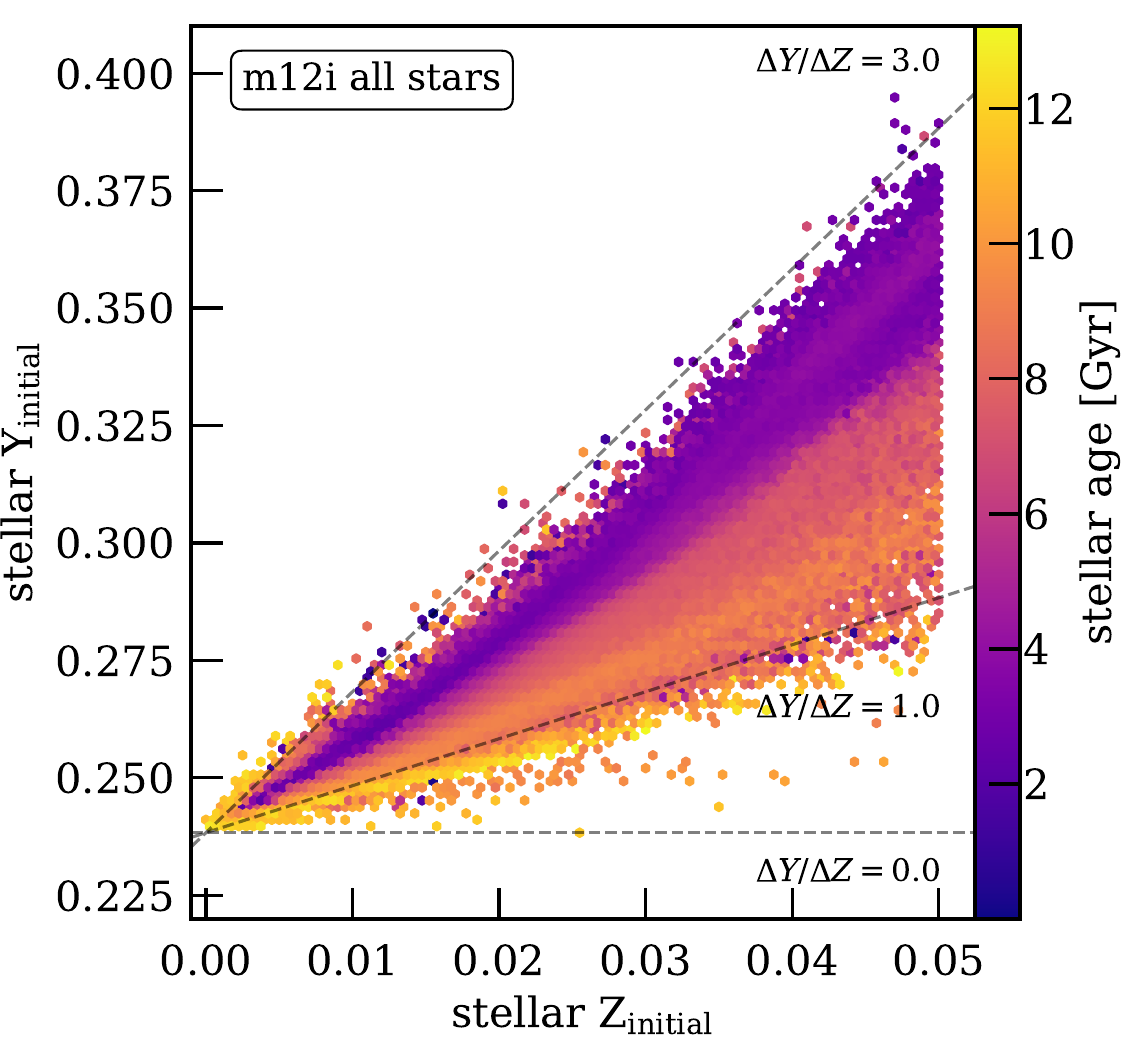} & \includegraphics[width = 0.32\linewidth]{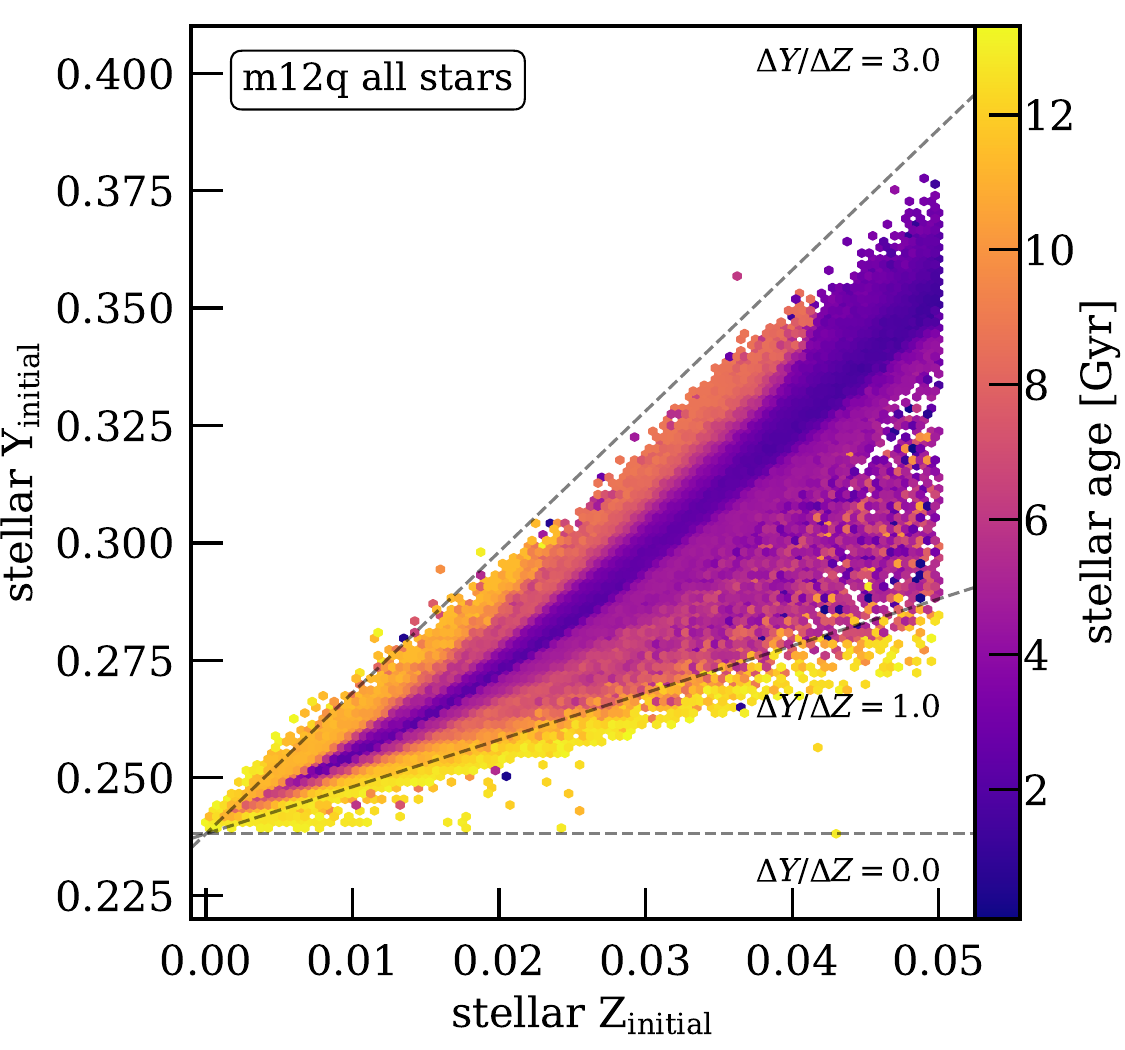} & 
\includegraphics[width = 0.32\linewidth]{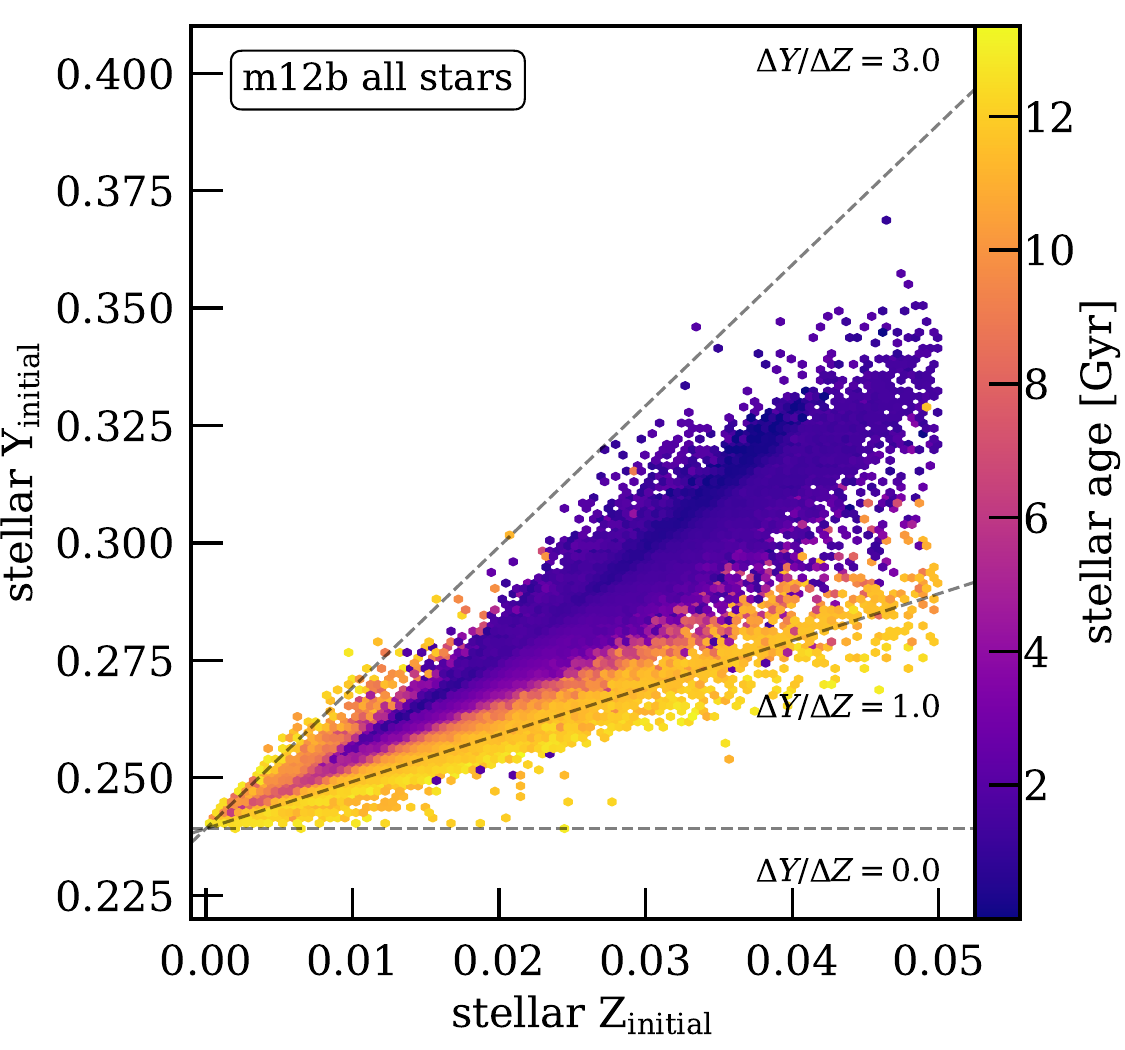} \\
\includegraphics[width = 0.32\linewidth]{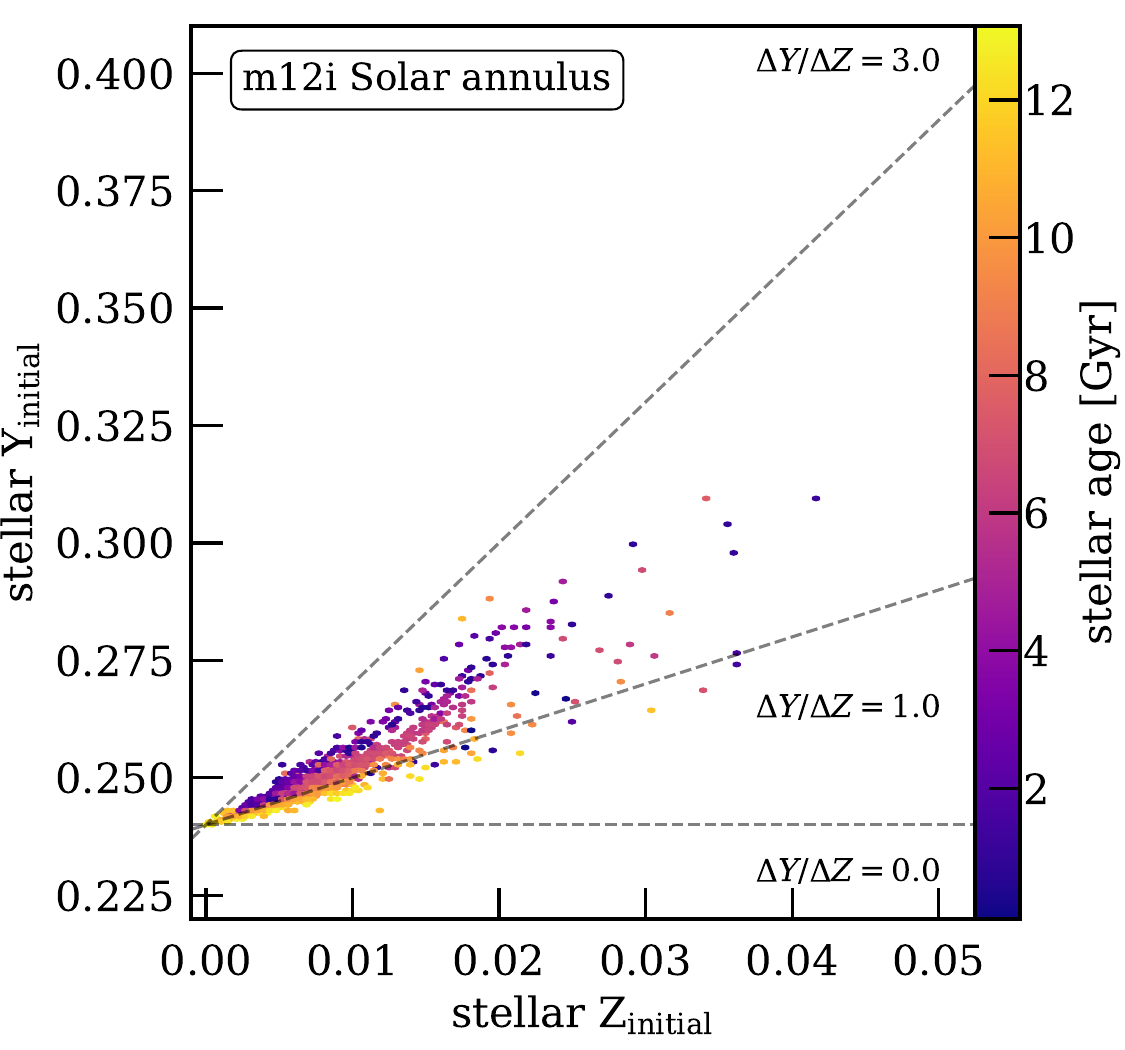} & \includegraphics[width = 0.32\linewidth]{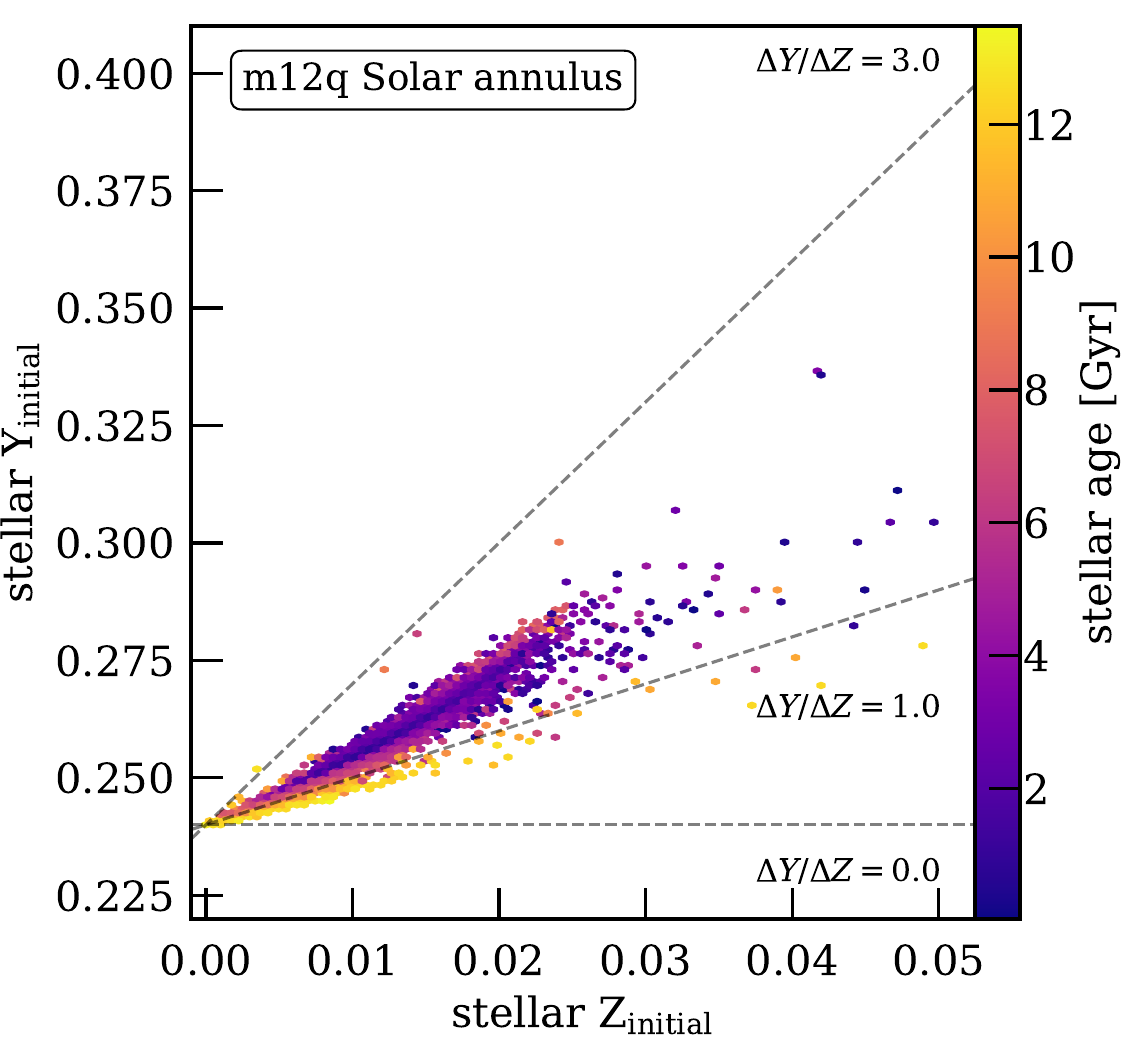} & 
\includegraphics[width = 0.32\linewidth]{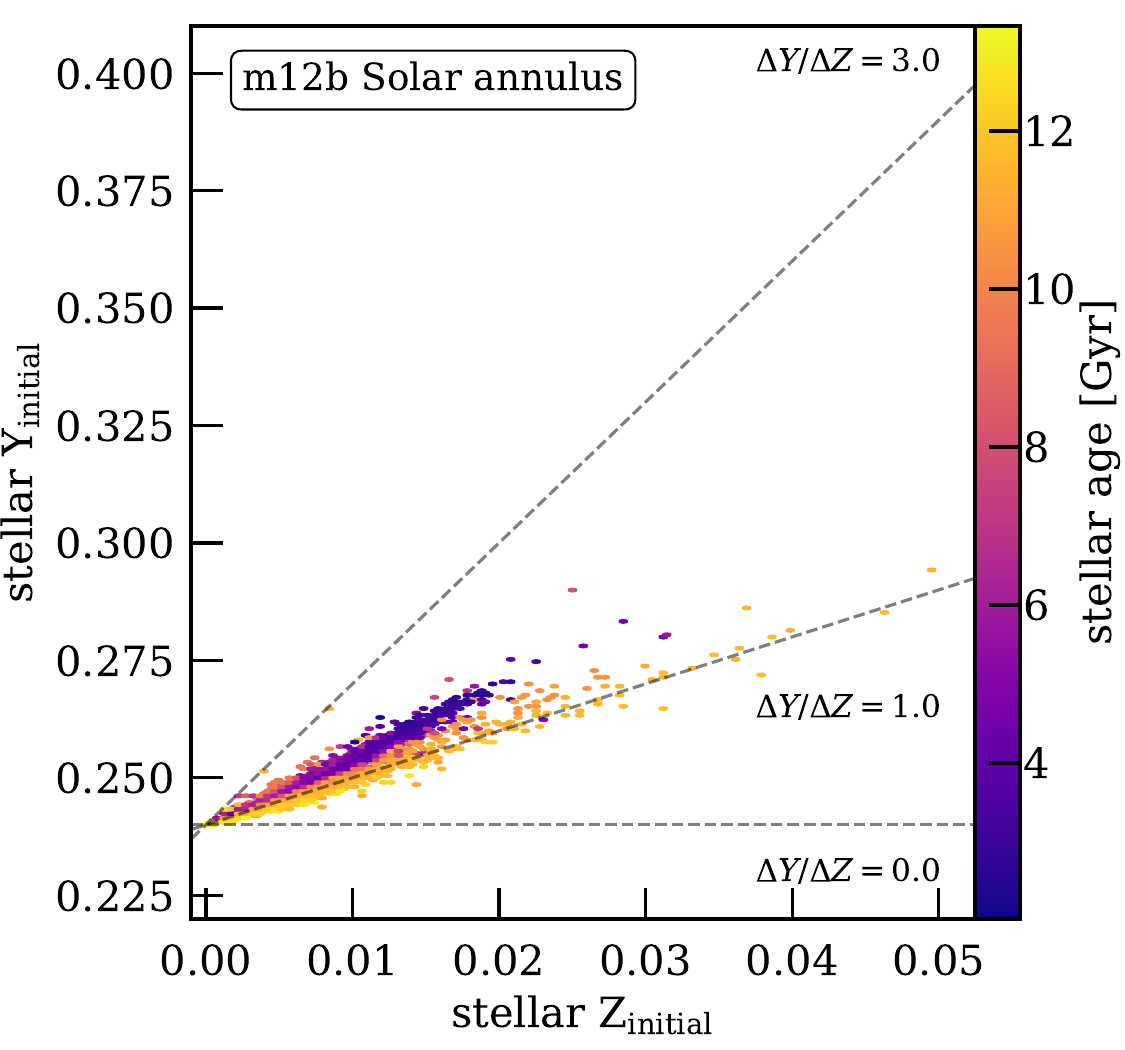} \\
\end{tabular}
\caption{
\textbf{Stellar $Y_{\rm initial}$ versus $Z_{\rm initial}$ for our three simulated MW-mass galaxies.} \textit{\underline{Top row:}} We plot abundances for all stars within a $15$ kpc sphere from the galactic center in each simulation, and exclude the minimal population ($< 5\%$) of star particles with $Z > 0.05$. \textit{\underline{Bottom row:}} Only stars in the Solar annulus in each simulation, with $Z \leq 0.05$. We color-code the plots by the average age at $z=0$ of the star particles in each hex bin. The dashed grey lines in each panel show different $\Delta Y$/$\Delta Z$ slopes to guide the eye, and the range from $0.0$ to $3.0$ brackets the full range suggested by fits to observed MW stars. The primary takeaway from each of the three simulations is that at $z=0$, stars lie along a wide range of $\Delta Y$/$\Delta Z$ slopes and do not conform to a specific, narrow range in slopes. However, the Solar annulus sample shows a smaller spread in $Y-Z$ space, and does not contain stars with highly super-Solar abundances (most of which lie in the bulge and inner galaxy in our simulations). Each simulation however, predicts a somewhat different variation of the slope with stellar age. Finally, we see that the youngest stars in each simulation ($\sim 0-2$ Gyr at present day) inhabit a narrower range in the Y-Z plane than older stars, although its precise slope varies across the three galaxies. 
}
\label{fig:main-stellar-Y-Z_plot}
\end{figure*}

We now discuss the co-evolution of stellar $Y$ and $Z$ in our simulated MW-mass galaxies. Figure~\ref{fig:main-stellar-Y-Z_plot} shows, for each of the three systems, hex-bin plots of the stellar initial helium mass fraction ($Y$) versus the stellar initial heavy metal mass fraction ($Z$), color-coded by stellar age in Gyr. 
For each galaxy, we consider both samples for this $Y-Z$ map: (a) all stars within $15$ kpc of the galaxy's center, and (b) only Solar annulus stars. 
To guide the eye, we overplot grey dashed lines showing different $\Delta Y$/$\Delta Z$ slopes of $0.0$, $1.0$, and $3.0$, similar to Figure 8 in \citet{silva-aguirre-17}.

Immediately, we see that there is no single, ``universal" $\Delta Y$/$\Delta Z$ relation that the stellar populations in our simulated galaxies follow, with the present-day stellar abundances spanning almost the entire range from below the $\Delta Y$/$\Delta Z = 1.0$ line up to the $\Delta Y$/$\Delta Z = 3.0$ line. There is also an age-dependence to the relation that various SSPs follow, with the youngest stars ($0-2$ Gyr) following a narrower trend in m12i and m12q but not so much in m12b; although the precise slope in the $Y-Z$ plane that the youngest stars follow varies across the three galaxies.    

Each of our three simulated galaxies shows a different spread in $\Delta Y$/$\Delta Z$ as a function of stellar age, which we attribute to the combined effects of the differences in their individual star formation histories (SFHs), coupled with varying nucleosynthetic yields as a function of progenitor SSP metallicity and the differing impacts of CCSN (more prompt) and AGB wind (more delayed) enrichment. In each galaxy, older stars ($\gtrsim 6$ Gyr) lie along somewhat different slopes in $Y-Z$ space than younger stars. However, that the precise slopes of the $Y-Z$ enrichment patterns for stars of different ages varies between the three galaxies, likely due to the differences in their SFHs at various points in cosmic time.

Additionally, the Solar annulus samples do not have stars going up to the extremely high $Y_{\rm initial}$ and $Z_{\rm initial}$ values that stars in the full sample within $15$ kpc show ($Y\approx0.38$, $Z\approx0.05$). This is because stars with the highest elemental abundances live in the bulges and inner disks in these galaxies, reflecting the radial metallicity gradients that have been well-studied in FIRE galaxies in previous studies, and agree well with observed gradients in the MW and other MW-mass galaxies \citep[][]{bellardini-22, graf-25a, graf-25b}. 

\begin{figure*}
\centering
\begin{tabular}{c c}
\includegraphics[width = 0.48 \linewidth]{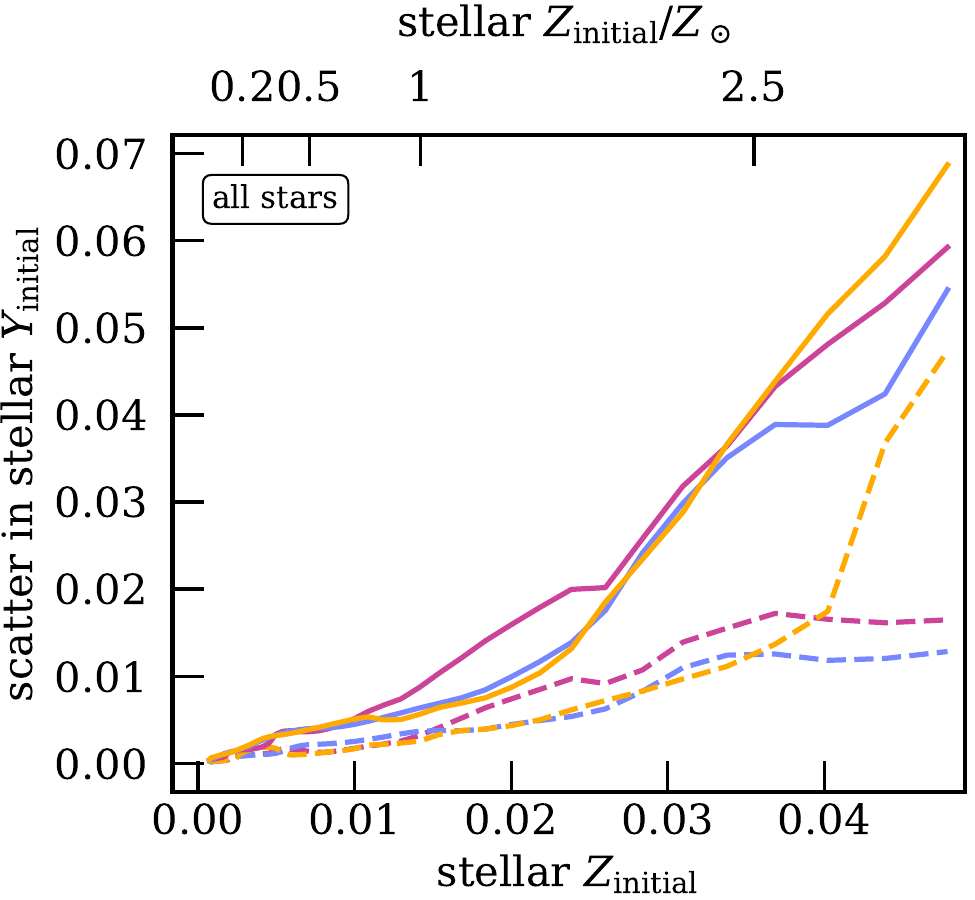} & 
\includegraphics[width = 0.48 \linewidth]{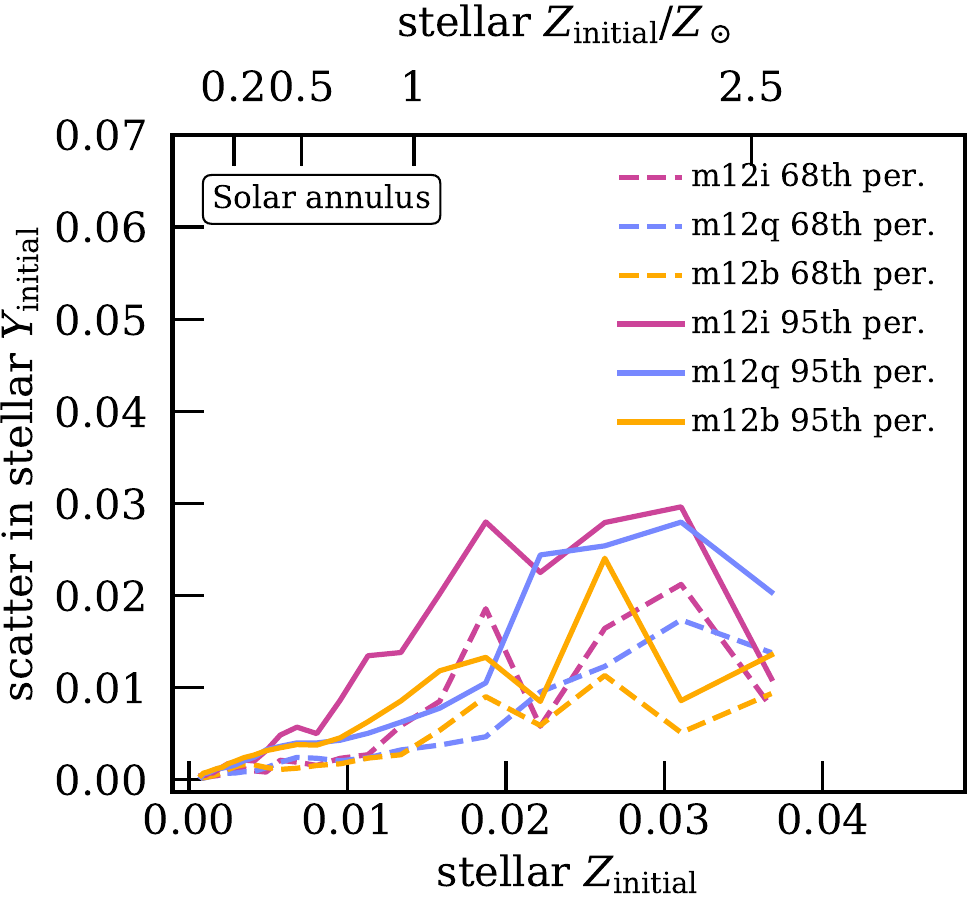}
\end{tabular} 
\caption{
\textbf{Spread in helium abundance as a function of metal abundance in our simulations.} Left panel shows the result for all stars within $15$ kpc, while the right panel shows only Solar annulus stars. For each galaxy, we show the scatter in stellar $Y_{\rm initial}$ in bins of $Z_{\rm initial}$, with dashed lines denoting the $68$th percentile range and solid lines showing $95$th percentile. Since Solar annulus stars do not exhibit extremely high metallicities that the ``all stars" samples do, we only show trend lines up to $Z\approx 0.035$. Scatter in $Y$ increases with increasing $Z$, and at $Z = Z_\odot$ is roughly $0.01-0.02$ ($95$th percentile), which is $\approx 4-8\%$ of $Y_\odot$. Solar annulus stars show marginally higher scatter in $Y_{\rm initial}$ compared to the sample of all stars, but qualitatively similar. Ultimately, $Y_{\rm initial}$ shows small-to-moderate (but not large) scatter as a function of $Z_{\rm initial}$ in our simulations, going up to a maximum of $\approx 20\%$ only for the most metal-rich stars in the galaxy, and topping out at a maximum of $\approx 10\%$ in the Solar neighborhood across all our simulated galaxies.
}
\label{fig:Y-scatter-vs-Z}
\end{figure*}

In Figure~\ref{fig:Y-scatter-vs-Z}, we examine the scatter in $Y_{\rm initial}$ as a function of $Z_{\rm initial}$ in each galaxy. The spread in $Y$ increases with increasing $Z$ for both the ``all stars" sample and the Solar annulus sample, ranging from $\approx 0.01-0.02$ at $Z_\odot$ (or $4-8\%$ of $Y_\odot$) to as high as $0.07$ at $Z\approx0.05$ (or $\gtrsim 20\%$ of $Y_\odot$). Thus, for a majority of stars in each galaxy, the $1\sigma$ scatter in $Y$ as a function of $Z$ is small-to-moderate, although the stars are spread out over a sizable part of the $Y-Z$ space in Figure~\ref{fig:main-stellar-Y-Z_plot}. Interestingly, the scatter in $Y$ for Solar annulus stars is comparable (or even marginally larger) than for the full sample of stars within $15$ kpc, for $Z\lesssim 0.03$. This is in part because of the smaller numbers of stars in the Solar annuli relative to the full galaxy-wide populations, and in part because they are somewhat more uniformly distributed in $Y-Z$ space.

\subsection{Best-fit $dY$/$dZ$ slopes and evolution over cosmic time}
\label{sec:3.x-dYdZ-slope}

For both samples in each simulated galaxy --- (a) all stars within $15$ kpc and (b) only Solar annulus stars --- we compute $\Delta Y$/$\Delta Z$ slopes. 
To measure best-fit slope values, we use non-linear least squares fitting via \texttt{scipy.optimize}. However, this method does not provide a meaningful metric for the scatter or variance in the slope. To obtain the $1\sigma$ scatter on the best-fit slopes, we instead compute the distribution of the number (or number density) of stars falling within different ``cones" in $Y-Z$ space defined by bins in $\Delta Y$/$\Delta Z \in [0.0, 4.0]$, anchored at primordial abundance values $Z=0.00, \, Y=0.24$. After manually verifying that these distributions are approximately Gaussian, we compute the $1\sigma$ of the distribution to give us the scatter in $\Delta Y$/$\Delta Z$. 

\begin{figure*}
\centering
\begin{tabular}{c}
\includegraphics[width = 0.7 \linewidth]{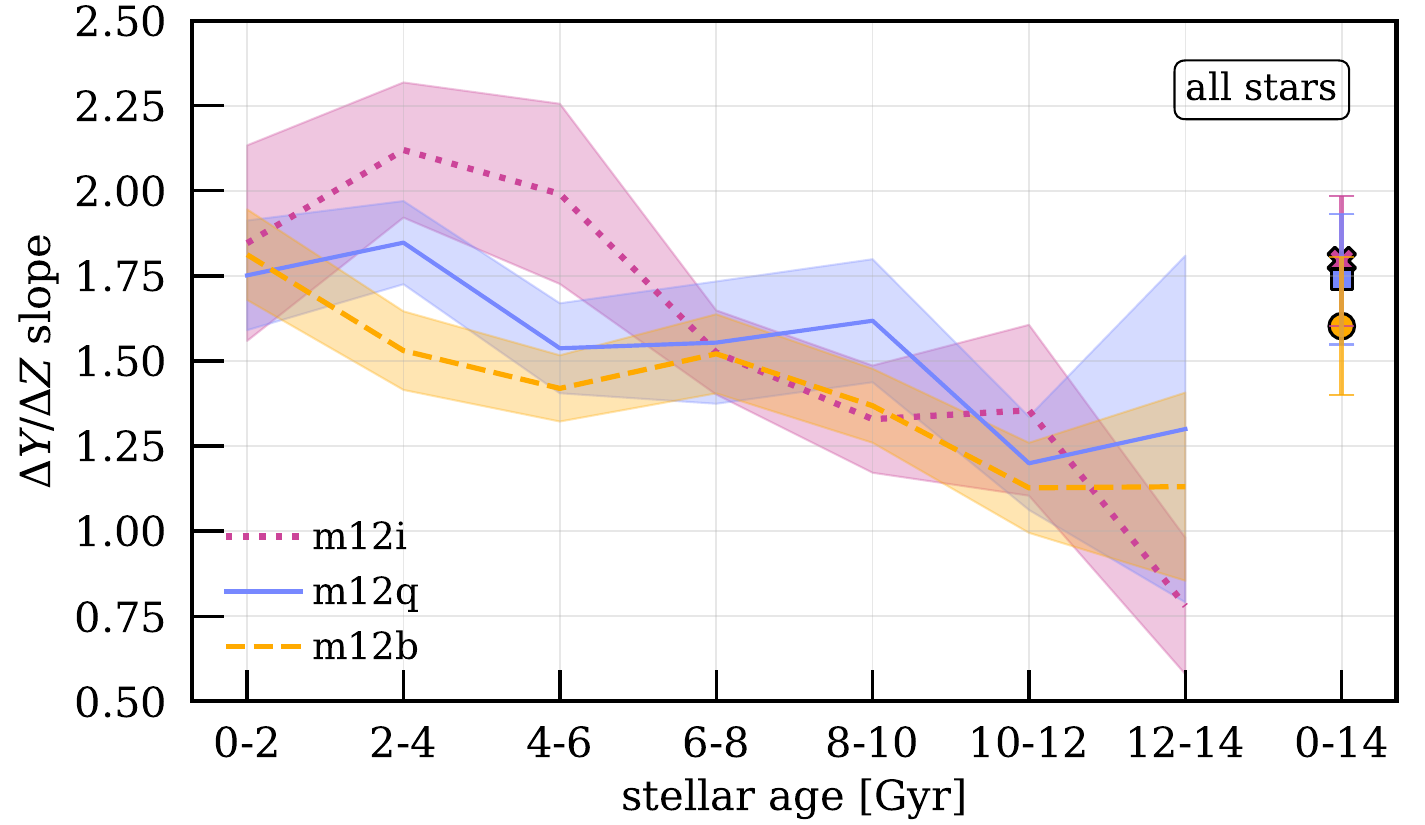} \\
\includegraphics[width = 0.7 \linewidth]{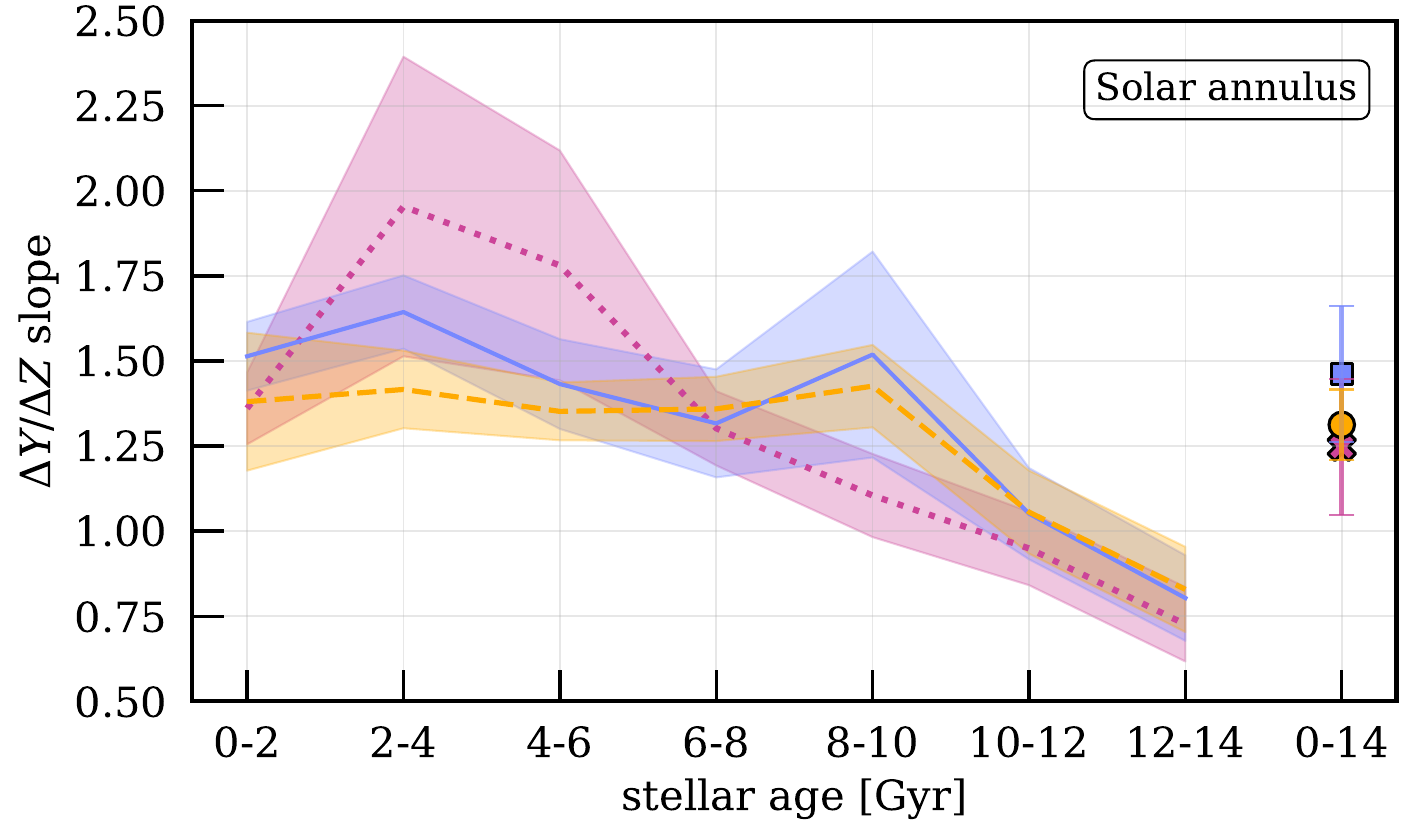}
\end{tabular} 
\caption{
\textbf{Variation in the helium enrichment relation over cosmic time.} For each galaxy (\texttt{m12i} in magenta, \texttt{m12q} in blue, \texttt{m12b} in gold) the curves show the best-fit $\Delta Y$/$\Delta Z$ value in bins of stellar age, and shaded regions show $1\sigma$ scatter. Scatter points show values computed using stars of all ages, corresponding to the values listed in Table~\ref{tab:slopes}. \textit{\underline{Top}}: for all stars within $15$ kpc from the galaxies' centers. \textit{\underline{Bottom}}: for only Solar annulus stars. 
Both panels demonstrate that to leading order $\Delta Y$/$\Delta Z$ increased as the galaxies built up over cosmic time, although galaxy-to-galaxy variations are present in individual stellar age bins depending on the specifics of each system's SFH. Broadly, $\Delta Y$/$\Delta Z$ ranges from $\approx 0.5-1.0$ for the oldest stars to $\approx 1.25-2.50$ for those younger than roughly $6$ Gyr. In each galaxy, the Solar annulus sample shows shallower slopes than the ``all stars" sample, by $\approx 0.25-0.50$; again, the specific value depends on the galaxy's formation history.
}
\label{fig:YZ-slope-vs-age}
\end{figure*}

Table~\ref{tab:slopes} shows the best-fit slope and $1\sigma$ scatter values for each galaxy, spanning $1.60-1.79$ with $\sigma \approx 0.20$ when considering all stars, and $1.25-1.46$ with $\sigma \approx 0.20$ for the Solar annulus. 
To demonstrate the evolution in the pattern of helium versus metal enrichment over time, Figure~\ref{fig:YZ-slope-vs-age} shows how $\Delta Y$/$\Delta Z$ varies as a function of stellar age in each simulation. 
To leading order, the oldest stars (which formed at the earliest times) show shallower slopes ($\approx 0.5-1.0$), with $\Delta Y$/$\Delta Z$ increasing towards younger ages. Solar annulus stars show shallower slopes overall compared to all stars within $15$ kpc, but in specific individual age bins they can be more comparable. Overall, the evolution of $\Delta Y$/$\Delta Z$ over time varies significantly amongst the three galaxies, highlighting the impact of their specific formation histories on the patterns of helium and metal enrichment. 

\subsection{Star formation histories and enrichment patterns of the FIRE-3 MW-mass galaxies}
\label{sec:3.x-SFH-comparison}

\begin{figure*}
\centering
\begin{tabular}{c c}
\includegraphics[width = 0.4785\linewidth]{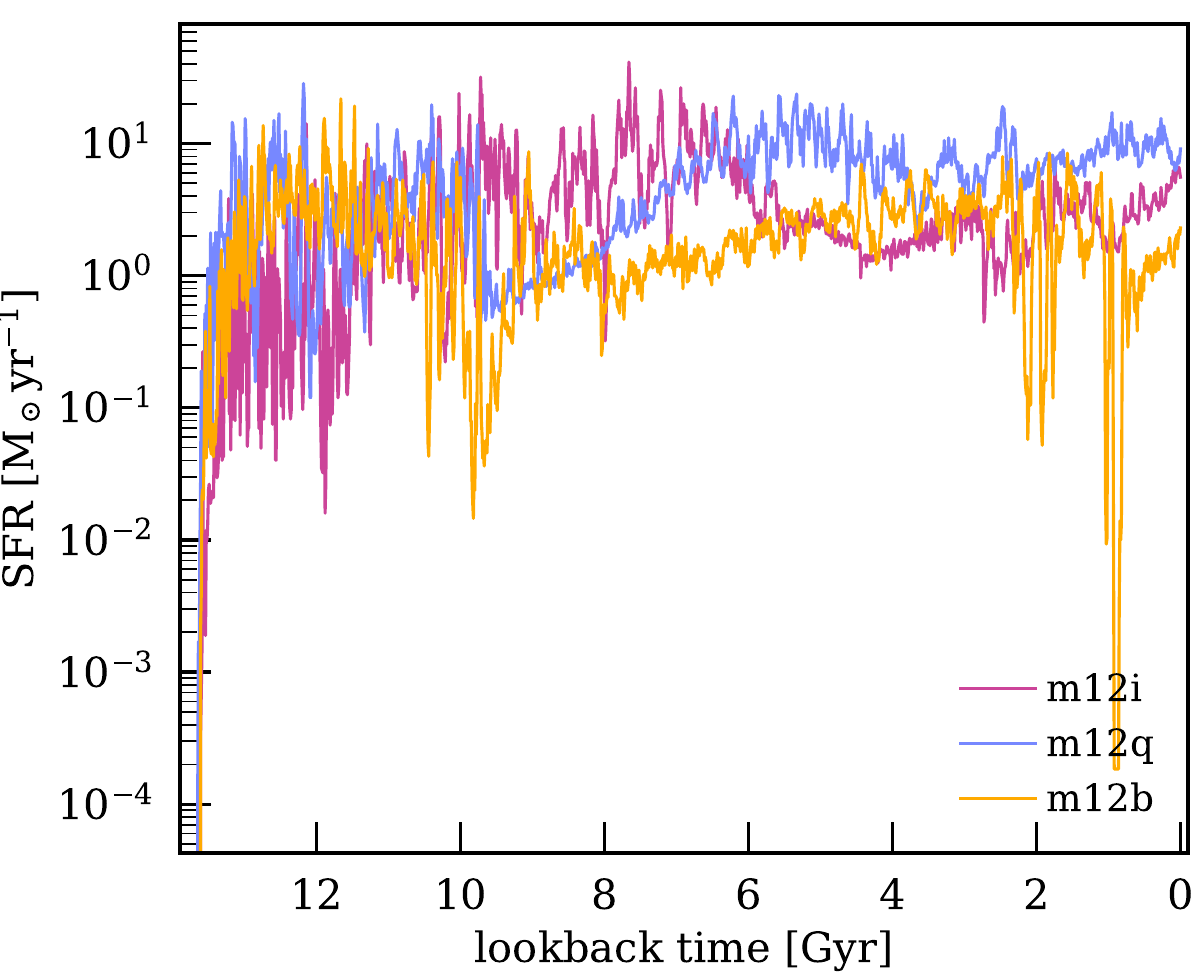} & \includegraphics[width = 0.4215\linewidth]{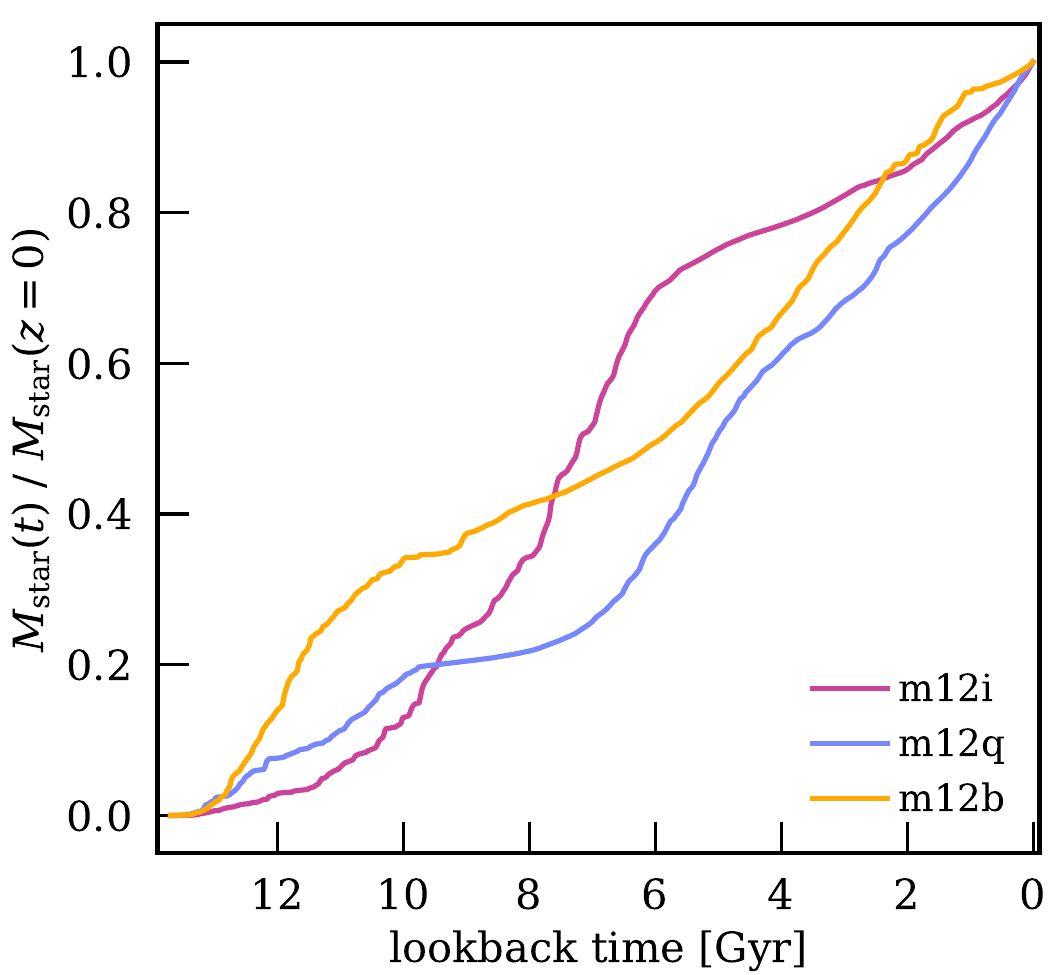}
\end{tabular}
\caption{
\textbf{Star formation histories of our three simulated galaxies.} \textit{\underline{Left:}} the star formation rate (SFR) for each galaxy in $\rm{M_\odot}\,\rm{yr}^{-1}$ as a function of lookback time (with $\approx 13.7$ Gyr ago being the Big Bang and $0$ Gyr being present day). \texttt{m12i} is in magenta, \texttt{m12q} in blue, and \texttt{m12b} in gold. We average the SFR over a running baseline of $0.01$ Gyr, or $10$ Myr, to highlight the inherently ``bursty" (stochastic) nature of star formation. Broadly, we show that the three galaxies have similar SFHs barring burstiness, with a few key differences: \texttt{m12q} and \texttt{m12b} show higher SFRs than \texttt{m12i} at lookback times $\gtrsim 11$ Gyr ago, \texttt{m12b} has the lowest SFR at $4 \lesssim t_{\rm lookback} \lesssim 8$ Gyr, \texttt{m12i} shows a dip in SFR for $3.5 \lesssim t_{\rm lookback} \lesssim 6$ Gyr, and most importantly \texttt{m12b} shows dramatic dips in SFR at lookback times of $\approx 2$ and $\approx 1$ Gyr, while \texttt{m12i} and \texttt{m12q} show relatively steady, slightly increasing, star formation from $\approx 2$ Gyr to present day. \textit{\underline{Right:}} the cumulative stellar mass growth histories of the galaxies relative to their final mass versus lookback time. The variations in SFR on the left map onto differences on the right. 
}
\label{fig:comparing-SFHs}
\end{figure*}

Since the present-day stellar elemental abundance patterns in a galaxy result from a combination of its star formation history (SFH) and nucleosynthetic enrichment from different stellar feedback channels, we next investigate the galaxy-wide SFHs of our simulated galaxies. The left panel of Figure~\ref{fig:comparing-SFHs} shows the archaeological star formation rates (SFRs) of our three MW-mass galaxies measured in $\rm{M_\odot}\, \rm{yr}^{-1}$ over all of cosmic time from the Big Bang ($t_{\rm lookback} \sim 13.7$ Gyr) to present day ($t_{\rm lookback} = 0$ Gyr). Here, ``archaeological" refers to the fact that we compute the SFHs for a galaxy from the formation masses of its star particles (within $15$ kpc) and their present-day ages. We compute SFR values averaged over a running baseline of $10$ Myr, to get an accurate sense of the inherent ``burstiness", or stochasticity. The right panel of Figure~\ref{fig:comparing-SFHs} shows the stellar mass growth of the three galaxies over cosmic time, in terms of their stellar mass at a given time relative to their present-day stellar mass.
 
As is expected for MW-mass galaxies, the SFRs for all three systems rise dramatically within the first $\approx 0.5-1$ Gyr of cosmic time to values of $\sim 0.1 - 10 \,\rm{M_\odot}\rm{yr}^{-1}$. From lookback times of $\approx 14-6$ Gyr, all our galaxies show significantly bursty SFHs with over $10\times$ variance above and below the average on short timescales. This is a well-known phenomenon from theoretical models of the formation of MW-mass galaxies (including FIRE) wherein at early cosmic times (until $z\sim1-2$) the galaxies have extremely bursty star formation and have not yet settled into well-ordered, rotation-supported disks \citep[][amongst others]{hopkins-23c, yu-23, mccluskey-24, semenov-24b}. Large variations in SFR on short timescales likely drive varying enrichment in helium and heavy metals from CCSN and O/B winds (shorter timescales) versus AGB winds (longer timescales), despite the fact that our simulations model metal mixing in the ISM. The lack of ordered disk-like rotation and the clumpy, patchy nature of the ISM \citep{bellardini-21, bellardini-22, graf-25b} in these galaxies at early cosmic times also leads to significant spatial (especially azimuthal) variation in enrichment across the galaxy, which also increases the spread in $Y-Z$ space for older stars at present day. This drives the larger spread in $\Delta Y$/$\Delta Z$ slopes across the three simulations for the oldest stars ($\gtrsim 12$ Gyr), as in Figure~\ref{fig:YZ-slope-vs-age}.

At later cosmic times ($t_{\rm lookback} \lesssim 6$ Gyr, and especially for $t_{\rm lookback} \sim 0-2$ Gyr), the galaxies show smoother SFHs with significantly less burstiness; the variance in SFR for each system is now less than $0.5$ dex around the average. This corresponds to the stage in their formation after their well-ordered disks have settled into thin disks \citep{hopkins-23c, yu-23, mccluskey-24, semenov-24b}. However, the average SFRs show large galaxy-to-galaxy variance (order of magnitude or higher) across the three simulations even if their individual SFHs are smoother, which persists to present day. Note that m12b shows a large drop in SFR $\approx 2$ Gyr ago and $\approx 1 Gyr$ ago. This likely drives the larger spread in \texttt{m12b}'s $\Delta Y$/$\Delta Z$ in its younger stars in Figure~\ref{fig:main-stellar-Y-Z_plot}, and is why \texttt{m12b} shows increasing $\Delta Y$/$\Delta Z$ slopes in the last $\approx 4$ Gyr in the top panel of Figure~\ref{fig:YZ-slope-vs-age}, while \texttt{m12i} and \texttt{m12q} do not. 

\begin{figure*}
\centering
\begin{tabular}{c c}
\includegraphics[width = 0.48\linewidth]{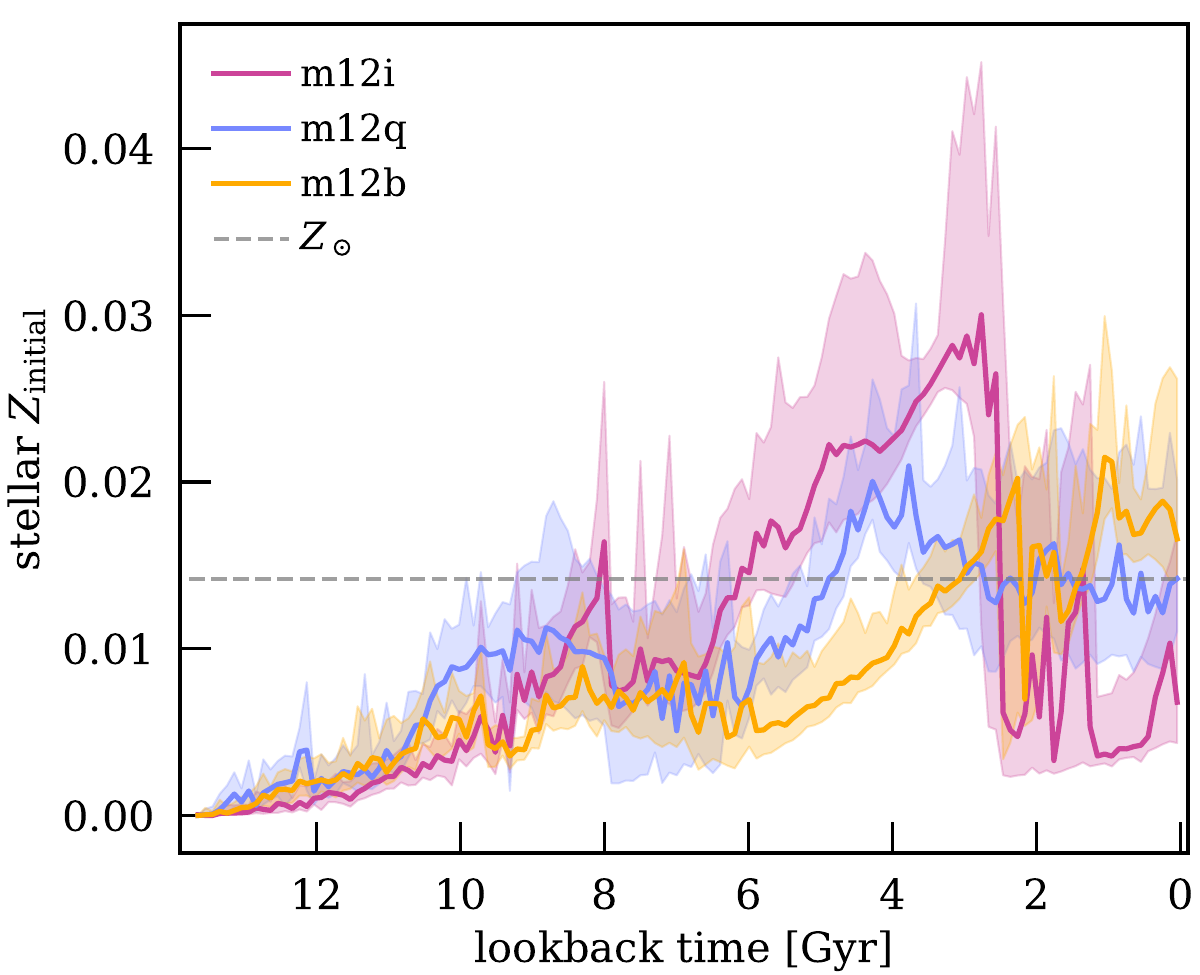} & \includegraphics[width = 0.48\linewidth]{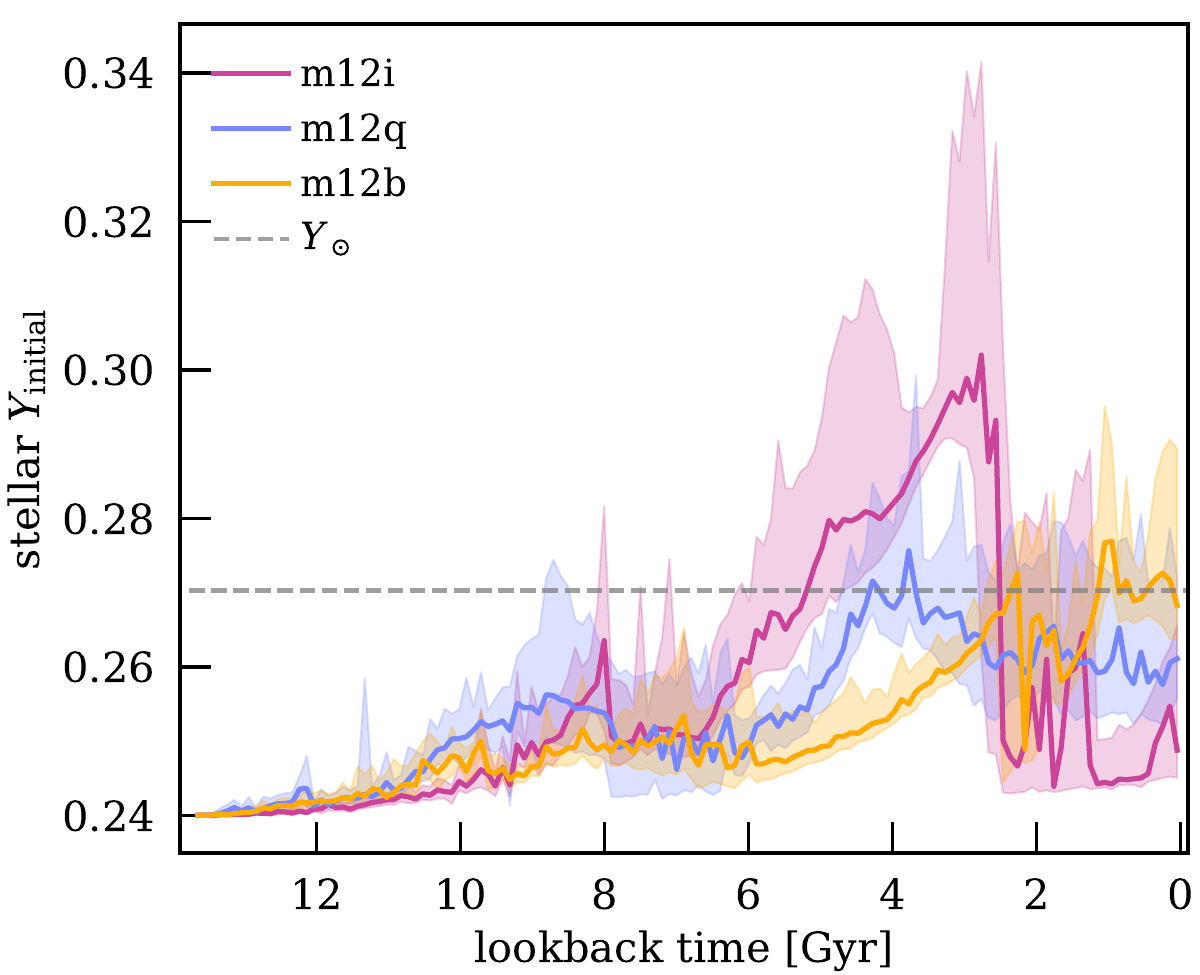} 
\end{tabular}
\caption{
\textbf{Stellar enrichment histories of our three simulated galaxies.} \textit{\underline{Left:}} the overall enrichment in the heavy metal mass fraction of stars $Z$, versus lookback time from the Big Bang ($\approx 13.7$ Gyr) to present day ($0$ Gyr). \textit{\underline{Right:}} enrichment in the helium mass fraction of stars $Y$ versus lookback time from the Big Bang to present day. The solid lines show the median value across all stars in the galaxy at the specific times, and the shaded regions show the $68\%$ scatter, with \texttt{m12i} in magenta, \texttt{m12q} in blue, and \texttt{m12b} in gold. In both panels, we compute median and $68\%$ values over a running $100$ Myr baseline. Overall, the three galaxies show increasing enrichment in stars and the ISM (since the composition of stars formed at a given time reflects the composition of the gas they formed out of) in both heavy metals and helium over time; however, there is significant variance in how quickly each galaxy enriches at given cosmic time (corresponding to the details of their star formation rates at that time). Also, we highlight the drops in metal/helium mass fraction that occur in each galaxy, with the most significant ones occurring at lookback times of $\approx 3$ Gyr (\texttt{m12i}), $\approx 4$ Gyr (\texttt{m12q}), and $\approx 2$ Gyr (\texttt{m12b}) --- these are the result of a significant merger with a more metal-poor lower-mass galaxy or a significant inflow of more pristine gas from the cosmic web, resulting in an influx of hydrogen and therefore lower $Z$ and $Y$.  
For an alternative view, see Appendix~\ref{sec:appendix-log-enrichment-histories} and Figure~\ref{fig:appendix-log-enrichment}.
}
\label{fig:comparing-enrichment-histories}
\end{figure*}

Differences in SFHs drive differences in elemental enrichment, since the elemental abundances of stars formed at a given time result from a convolution of the star formation rate and feedback/enrichment patterns. In Figure~\ref{fig:comparing-enrichment-histories} we show separately the evolution over time of stellar $Z_{\rm initial}$ and stellar $Y_{\rm initial}$ (see Appendix Figure~\ref{fig:appendix-log-enrichment} for an alternative depiction). To leading order, all three simulated galaxies show an increase in average $Z_{\rm initial}$ and $Y_{\rm initial}$ from the Big Bang ($t_{\rm lookback} \sim 13.7$ Gyr) to present day ($t_{\rm lookback} = 0$ Gyr) as is expected. However, they also show significant time-dependent deviations from the average trend and galaxy-to-galaxy variance. As with the SFHs, the enrichment histories have an inherent level of stochasticity, owing to the bursty nature of star formation coupled with the stochasticity of the stellar feedback processes (especially supernovae) that sources enrichment.

To leading order, the enrichment in helium tracks the enrichment in heavy metals because they are both sourced from CCSN on shorter timescales after a burst of star formation ($\lesssim 400$ Myr), as shown in Figures~\ref{fig:yields-vs-age} and~\ref{fig:channels-per-element}. However, Figure~\ref{fig:comparing-enrichment-histories} shows higher-order differences in the exact enrichment of $Z$ and $Y$, since later-time AGB winds also contribute to helium enrichment, especially for more metal-rich stellar populations, and later-time white dwarf supernovae contribute only metals but no helium. We also see in both panels sudden drops in $Z$ or $Y$ values at specific times; these are the result of an influx of pristine gas into the ISM (from either a gas-rich merger of a lower-mass galaxy or gas accretion from the cosmic web) and drive up the hydrogen fraction ($X$) in the ISM, thus lowering the $Y$ and $Z$ abundances of the stars that form out of the ISM gas for some time thereafter. These influxes of pristine gas also contribute to the evolution in $Y-Z$ space since they reset the abundances of stars forming afterwards to lower values.

Figure~\ref{fig:comparing-enrichment-histories} also shows that the most significant drop in $Z$ and $Y$ occurs in \texttt{m12i} at $t_{\rm lookback} \sim 2-3$ Gyr, and occurs after the galaxy's most significant enrichment period during $t_{\rm lookback} \sim 3-7$ Gyr. The significant enrichment of both $Y$ and $Z$ relative to \texttt{m12q} and \texttt{m12b} is because \texttt{m12i} assembles the largest fraction of its present-day stellar mass compared to the other two systems during that time period ($\sim 3-7$ Gyr ago, right panel of Figure~\ref{fig:comparing-SFHs}), and this increase in stellar mass and the corresponding increase in stellar feedback drive up its enrichment. \texttt{m12q} and \texttt{m12b} also show periods of increasing enrichment during $t_{\rm lookback} \sim 4-7$ Gyr and $t_{\rm lookback} \sim 2-6$ Gyr respectively, but at a steadier, slower rate than m12i because of their lower average SFRs during that period (as seen in the left panel of Figure~\ref{fig:comparing-SFHs}).

Ultimately, the elemental abundance patterns of a star formed at a specific time in a galaxy's history are due to the combined, cumulative effects of: (a) the galaxy's SFH up to that time, (b) the relative enrichment from different feedback channels operating on different timescales, (c) turbulent metal mixing in the ISM, and (d) metal loss due to feedback-driven galactic winds (outflows) as well as the influx of more pristine gas (inflows) from the large-scale cosmic web. This complex interplay of chaotic processes makes it difficult to deterministically predict the exact abundance patterns of a stellar population or the exact evolution in $\Delta Y$/$\Delta Z$. However, broadly speaking we argue that the different yields from CCSN versus stellar winds are key for determining the time-evolution of $\Delta Y$/$\Delta Z$. At early times when the galaxy is more metal-poor, CCSN dominate both metal and helium production, and produce proportionally lower helium-to-metals than winds (left column of Figure~\ref{fig:yields-vs-age} and~\ref{fig:channels-per-element}). Thus, the early-time $\Delta Y$/$\Delta Z$ values in each galaxy are lower, around $0.5-1.0$, as Figure~\ref{fig:YZ-slope-vs-age} shows. As time progresses and the galaxy gets more enriched on average, stellar winds begin to contribute more (AGB winds especially create late-time enrichment), and produce proportionally higher helium-to-metals than supernovae (middle and right columns of Figure~\ref{fig:yields-vs-age} and~\ref{fig:channels-per-element}). Therefore, the $\Delta Y$/$\Delta Z$ values in each galaxy increase over time on average (Figure~\ref{fig:YZ-slope-vs-age}). Short-timescale variations in a galaxy's SFR combined with the stochastic effects of supernovae and turbulent metal mixing create the higher-order variations relative to the average trend that we see in the evolution of $\Delta Y$/$\Delta Z$ over time. Finally, the difference between the Solar annulus samples and the ``all stars" samples can be attributed to the fact that stars born at specific radii can be redistributed over time due to dynamical effects (also called ``radial migration"), so the stars in the Solar annulus at present day were not necessarily born in that region. Previous studies have shown that FIRE galaxies demonstrate emergent radial redistribution of stars on a level comparable to that of the MW, giving confidence that our results for the Solar annulus relative to the whole galaxy are physical and not simple a numerical artifact of our simulations.

\section{Conclusions} 
\label{sec:4-conclusions}

\subsection{Summary of key results}
\label{sec:4.x-summary}

Using three cosmological baryonic zoom-in simulations of MW-mass galaxies --- \texttt{m12i, m12q, m12b} --- from the FIRE-3 project, we investigated the patterns of stellar $Y_{\rm initial}$ and $Z_{\rm initial}$ in their stars at present day, also called the ``chemical enrichment law".
\begin{itemize}
    \item According to the feedback rate and nucleosynthetic yield models we assume in FIRE-3 (Figures~\ref{fig:yields-vs-age} and~\ref{fig:channels-per-element}), the total heavy metal yield is dominated by core-collapse supernovae (CCSN) at all progenitor SSP metallicities, and enrichment occurs within $\approx 250$ Myr after a burst of star formation. CCSN are a significant contributor to the helium yield within $\approx 400$ Myr after star formation, especially for lower progenitor SSP metallicities, but late-time AGB winds also contribute to helium enrichment, and dominate over CCSN for more metal-rich SSPs. 
    \item Our simulated galaxies show a wide spread in $Y-Z$ space at present day, spanning a range of $\Delta Y$/$\Delta Z$ values of $\approx 0.5 - 2.5$ (Figures~\ref{fig:main-stellar-Y-Z_plot} and~\ref{fig:YZ-slope-vs-age}, and Table~\ref{tab:slopes}). 
    Older stars ($\gtrsim 6-8$ Gyr) show shallower slopes than stars younger than $\sim 6$ Gyr across all three simulations, although the precise variations in the trend versus time are susceptible to galaxy-to-galaxy variations in star formation histories. 
    \item \textit{No universal ``enrichment law"}: our simulations show that helium-versus-metal enrichment patterns for stars of different ages do not follow a single $\Delta Y$/$\Delta Z$ enrichment law (Figure~\ref{fig:YZ-slope-vs-age}, Table~\ref{tab:slopes}). For Solar annulus stars, they predict best-fit $\Delta Y$/$\Delta Z$ values of $1.25-1.46$ with $\sigma \approx 0.2$. When considering all stars within $15$ kpc, they predict somewhat higher $\Delta Y$/$\Delta Z = 1.60 - 1.79$ with $\sigma \approx 0.2$. However, the helium enrichment relation varies significantly over the course of the galaxies' formation histories, with $\Delta Y$/$\Delta Z \approx 0.5-1.0$ at early cosmic times, going up to a maximum of $\Delta Y$/$\Delta Z \gtrsim 2.0$ roughly $2-6$ Gyr ago.
    \item The star formation histories (SFHs) of our three simulated galaxies demonstrate significant ``burstiness" (stochasticity), as is expected for MW-mass galaxies especially at early times. They also show galaxy-to-galaxy variance in their average SFHs at all cosmic times (Figure~\ref{fig:comparing-SFHs}). These differences in SFHs, coupled with differential enrichment from stellar winds versus core-collapse supernovae at different progenitor metallicities, maps on to the differences in their helium enrichment and metal enrichment over time (Figure~\ref{fig:comparing-enrichment-histories}), and ultimately on to differences in $Y-Z$ space and the evolution of their $\Delta Y$/$\Delta Z$ slopes.
\end{itemize}

\subsection{Discussion}
\label{sec:4.x-discussion}

\begin{table}
\centering
\caption{
Quantifying $Y$ versus $Z$ enrichment patterns in our FIRE-3 MW-mass simulations
}
\vspace{-2 mm}
\addtolength{\tabcolsep}{3 pt}
\begin{tabular}{ccc}
\hline
\hline
\vspace{1mm}
Simulation & $\Delta Y$/$\Delta Z$ & $\Delta Y$/$\Delta Z$\\
name & (all stars in $15$ kpc) & (Solar annulus)\\
\hline
\texttt{m12q} & $1.74 \pm 0.18$ & $1.46 \pm 0.19$ \\
\texttt{m12i} & $1.79 \pm 0.19$ & $1.25 \pm 0.20$ \\
\texttt{m12b} & $1.60 \pm 0.20$ & $1.31 \pm 0.10$ \\
\hline
\hline
\end{tabular}
\addtolength{\tabcolsep}{-3 pt}
\tablecomments{
For each galaxy, we show the $\Delta Y$/$\Delta Z$ slope at $z=0$ computed using (a) all stars within $15$ kpc and (b) only stars within the Solar annulus ($7.5\,\rm{kpc} \leq \rm{r} \leq 8.5\,\rm{kpc}$, $\rm{z} \leq |1|\,\rm{kpc}$). Best-fit slope and scatter values are as per the technique described in Section~\ref{sec:3.x-dYdZ-slope}. The Solar annulus slopes are systematically shallower than the Galaxy-wide ones, as Figure~\ref{fig:YZ-slope-vs-age} also shows.
}
\label{tab:slopes}
\end{table}

First, we address some caveats regarding our analysis and its interpretation. In the FIRE-3 model (as is standard for all cosmological zoom-in simulations of galaxy formation), we make sub-grid assumptions for mass/energy/momentum/metal loss rates from stellar feedback channels and the nucleosynthetic yields. Although our choices are physically motivated based on stellar evolution and supernova models from the literature rather than fine-tuned to produce specific outcomes, there exists a fair amount of uncertainty for feedback rates and yields in the literature, which can impact on the emergent SFHs and elemental abundances of simulated galaxies \citep[as in][]{hopkins-18b, agertz-20, vogelsberger-20, muley-21, gandhi-22, sales-22, andersson-26, kobayashi-26}. However, our goal in this work is not to make definitive claims about which yield models are better or worse; the takeaway instead is that even with the choice of one fixed set of feedback rates and yields, the self-consistent modelling of SFH and enrichment in our simulations leads to a spread in the resultant $Y_{\rm initial}$-$Z_{\rm initial}$ space, rather than a single "universal" trend .

Additionally, as discussed in Section~\ref{sec:2.x-fire}, each star particle in FIRE represents not an individual star but rather a mono-age simple stellar population following a \citet{kroupa-01} IMF. When comparing our results to observations of that are used to constrain $\Delta Y$/$\Delta Z$, it is therefore not an apples-to-apples comparison since the observations are of individually resolved stars in the MW. For this particular work, this is not an issue because rather than comparing the compositions of individual stars, we are concerned more with the overall trends and variations in the $Y$-$Z$ space. Since the star particles in FIRE inherit the elemental compositions of the gas elements they form out of, they are still representative of the spatial and temporal variations in $Y$ and $Z$ enrichment that we should expect from MW-mass galaxies. For the question of what individual stellar compositions look like in the simulations, future work with synthetic star-by-star observations using pipelines like \textsc{Ananke} \citep{sanderson-20, thob-24} will be useful. One additional caveat is that the elemental abundances of star particles in FIRE do not change after formation, despite accounting for metal loss via stellar feedback. This remains important context, and is also why we only discuss the initial $Y$ and $Z$ abundances for stars in our simulated galaxies.

We note that the simulated galaxies in the FIRE project were chosen to be in the expected range for the total mass of the MW, but were not designed to be perfect MW analogs \citep{wetzel-23}. For example, they do not trace exactly the MW's formation history and do not all include proper analogs for the LMC or a GSE merger. Given the recent observational evidence that the real MW is somewhat of an outlier system that formed unusually early for a galaxy of its mass and accreted its satellite population proportionally earlier \citep[][]{belokurov-22, conroy-22, rix-22, mao-24, semenov-24a, mccluskey-25}, we do not expect any of our FIRE-3 galaxies to exactly reproduce the SFH and enrichment history of the MW. Therefore, our results for enrichment patterns and the spread in $\Delta Y$/$\Delta Z$ should not be understood as exactly representing the actual stellar abundances in the MW. Rather, we present them as probing the range of possible trends in helium versus metal enrichment in MW-mass galaxies, given the astrophysical variance in formation histories expected for galaxies at similar present-day masses (due to baryonic effects, halo-to-halo scatter, and the inherent stochasticity in star formation and feedback processes). However, \citet{steel-26} showed that \texttt{m12q} is the earliest forming FIRE MW-mass galaxy, likely pushing closer to the actual MW's formation history. Thus, out of our three simulations, the results for \texttt{m12q} are likely the closest to those of the real MW, even if not a perfect analog. Our broader point that the peculiarities of a galaxy's SFH and enrichment history affect the variations we see in its present-day $Y$-$Z$ space still holds true.

To our knowledge, \citet{vincenzo-19} is the only other study that examined the enrichment patterns of helium and metals (with $\Delta Y$/$\Delta Z$ in mind) in cosmological simulations of MW-mass galaxies. Their simulations differ in a number of key ways from ours. They relied on MW-mass galaxies chosen from a large-volume (or ``big box") cosmological simulation while we used the zoom-in technique on MW-mass galaxies from a cosmological volume; as a result our simulations have $\approx 1000\times$ higher mass and spatial resolution. Their simulations also utilized a different code and different prescriptions for nucleosynthetic yields as well as how stellar feedback couples to the ISM. Despite these differences, our main results agree well with those from \citet{vincenzo-19}: that $\Delta Y$/$\Delta Z$ is not fixed within a galaxy, evolves over cosmic time, and depends on the details of the galaxy's formation history. The agreement between our simulations and theirs is encouraging, and indicates that the results are physical rather than due to choices of sub-grid feedback prescriptions or numerical techniques. We do, however, predict higher scatter in $Y$ as a function of $Z$ and slightly higher $\Delta Y$/$\Delta Z$ slopes over time than \citet{vincenzo-19} do; compare our Figure~\ref{fig:Y-scatter-vs-Z} to their Figure 11, and our Figure~\ref{fig:YZ-slope-vs-age} to their Figure 6. These higher-order differences are likely due to the different SFHs of our galaxies compared to theirs, and the differences in our choices of feedback rate and nucleosynthetic yields. Overall though, our work shows good agreement with their results.  

Our key result that there is no single chemical enrichment law that applies to all stars in the MW is in agreement with studies that conducted observational inferences of $\Delta Y$/$\Delta Z \approx 1.0-3.0$ \citep[such as][]{balser-06, casagrande-07, serenelli-10, verma-14, corsaro-15, silva-aguirre-17, verma-19, saunders-23}. These studies often point to small sample sizes of asteroseismically analyzed stars or systematics in modelling techniques as the reason for the wide spread in $\Delta Y$/$\Delta Z$. However, we show that this diversity in the $Y-Z$ plane is likely not just due to selection effects or uncertainties, but rather represents a genuine, physical spread in the enrichment patterns of He relative to heavy elements, driven by spatial/temporal variations in the SFH of the MW as well as by the competing effects of feedback from CCSN (more prompt) and AGB stellar winds (more delayed). Our simulations self-consistently model the interplay between dark matter, gas, and stars in MW-mass galaxies while tracking star formation, stellar feedback, and elemental enrichment over time, showing that there exists an emergent spread in $\Delta Y$/$\Delta Z$ that cannot be encapsulated within a single, fixed value of a ``chemical enrichment law", but rather evolves over cosmic time. 

However, the $Y-Z$ space still represents an organized mapping between a galaxy's formation history and its elemental abundances, and thus it remains extremely valuable to measure $Y_{\rm initial}$, $Z_{\rm initial}$, and age from asteroseismology and spectroscopy with accurate stellar models. Obtaining accurate measurements of helium and metal abundances for large samples of MW stars could potentially help add constraints to our Galaxy's formation history. Most observational efforts in asteroseismology and constraining $\Delta Y$/$\Delta Z$ have so far been limited to nearby stars in the Solar neighborhood. With upcoming data from the \textit{Roman} \citep{roman-23, roman-25} and PLATO \citep{rauer-25} missions, we will have access to observations of $\sim$ billions of stars in the Galactic bulge, inner disk, and stellar halo. Our results for how $\Delta Y$/$\Delta Z$ differ between the Solar annulus and the Galaxy-wide samples help inform the need for a more careful treatment of the initial helium abundance in stellar models in the era of \textit{Roman} and PLATO.

For a discussion of how choices of initial helium abundance map onto systematic uncertainties in asteroseismic inferences of stellar age and other parameters, see \citet{lebreton-14}. They investigate how the ``mass-helium degeneracy" in modelling --- the well-known degeneracy where higher $Y_{\rm initial}$ requires lower mass in models \citep[e.g.,][]{verma-22} --- can affect inferences of fundamental stellar properties like ages, masses, and radii. In particular, they show that assuming a fixed value for $Y_{\rm initial}$ (from a $\Delta Y$/$\Delta Z$ scaling relation, for example) leads to a large systematic scatter in age of $\gtrsim 3-4$ Gyr, when using a star's effective temperature ($T_{\rm eff}$), luminosity ($L$), and iron metallicity ([Fe/H]) measurements as inputs. Asteroseismic modelling that allows $Y_{\rm initial}$ to be a free parameter results in smaller uncertainties on age ($\approx 15-20\%$), assuming one can obtain precise seismic measurements of the star's pulsation frequencies. However, the mass-helium degeneracy still significantly impacts the inference of a star's mass even in the case of asteroseismology. 

More recently, \citet{nsamba-21} used $35$ low-mass, Solar-type stars from the `LEGACY' sample with multi-year \textit{Kepler} photometry to investigate the systematic effects of the treatment of initial He abundance in stellar models on inferred stellar parameters like mass, radius, mean density, and age. They found that stellar model grids using higher values of $\Delta Y$/$\Delta Z$ led to lower inferred masses and radii compared to those with lower $\Delta Y$/$\Delta Z$, but with small systematic uncertainties of a few per cent. Setting initial $Y$ as a free parameter in the stellar modelling led to smaller masses and radii than with fixed $\Delta Y$/$\Delta Z$. Strikingly, while they only found small systematic differences of a few per cent when inferring masses, radii, and mean densities between the models with $Y$ as a free parameter and those with fixed enrichment laws, they found large systematics of $\sim 30 \%$ for age. Given the spread in $\Delta Y$/$\Delta Z$ we observe in our simulations, the \citet{nsamba-21} results suggest that asteroseismic ages of MW stars include non-negligible systematic uncertainties from the assumptions of initial He abundance. 

Galactic archaeology has been an incredibly successful field, helping us uncover the formation and accretion history of the MW, the presence of the ancient Gaia-Sausage-Enceladus merger, the spin-up and settling of the MW's stellar disk over time, and more \citep[see the review by][]{deason-24}. This revolution in our understanding of our own Galaxy has relied heavily on the use of stellar evolution models to measure ages, masses, and elemental abundances of MW stars, and will be further improved by the wealth of photometric and asteroseismic data we will receive from upcoming missions such as \textit{Roman} and PLATO. Understanding the assumptions and sources of uncertainty in our stellar models is of paramount importance in this golden age of MW science, as we have shown with this analysis of the impact of helium abundance assumptions.

\section{Software and third-party data repository citations} 
\label{sec:cite}

\textit{\large{Data availability:}} 
A public version of the \textsc{Gizmo} code is available at \url{http://www.tapir.caltech.edu/~phopkins/Site/GIZMO.html}. Although the FIRE-3 simulations are not publicly accessible, the FIRE-2 versions are available at \url{https://flathub.flatironinstitute.org/fire} \citep{wetzel-23, wetzel-25}. 

\software{
        arXiv pre-print service\footnote[4]{\url{https://www.arxiv.org}},
        NASA ADS\footnote[5]{\url{https://ui.adsabs.harvard.edu}},
        Gizmo Analysis\footnote[6]{\url{https://bitbucket.org/awetzel/gizmo\_analysis}} \citep{wetzel-20-gizmo},
        Halo Analysis\footnote[7]{\url{https://bitbucket.org/awetzel/halo\_analysis}} \citep{wetzel-20-halo},
        ipython \citep{ipython-07},
        matplotlib \citep{matplotlib-07}, 
        numpy \citep{numpy-20},
        scipy \citep{scipy-20},
}

\begin{acknowledgments}

PJG thanks Ebraheem Farag, Marla Geha, and Frank van den Bosch for useful discussions during the preparation of this manuscript. EPB received supported from NSF award No.\ 2606207. AW received support from NSF, via CAREER award AST-2045928 and grant AST-2107772. JB acknowledges support from NASA grant number 80NSSC24K1227. IT received support from NSF AAPF 2602617.  

\end{acknowledgments}

\begin{contribution}

PJG ran simulations, led the analysis, and led the writing of this manuscript. EPB played a crucial role in the original motivation of this work and provided insights on stellar evolution. AW contributed to running simulations and provided valuable insight into stellar feedback processes relevant to our study. JB gave useful guidance on the broader context of comparing to the Milky Way's formation history. SB provided insights into stellar models and gave feedback on the presentation of ideas. IT contributed to and provided feedback on the discussion of enrichment from different feedback channels. FJM provided feedback on relating our results to galaxy formation theory. All authors contributed to the final polishing of the manuscript.


\end{contribution}

%



\appendix

\section{Stellar enrichment histories of our three simulated galaxies}
\label{sec:appendix-log-enrichment-histories}

\begin{figure}
\centering
\begin{tabular}{c}
\includegraphics[width = 0.5\linewidth]{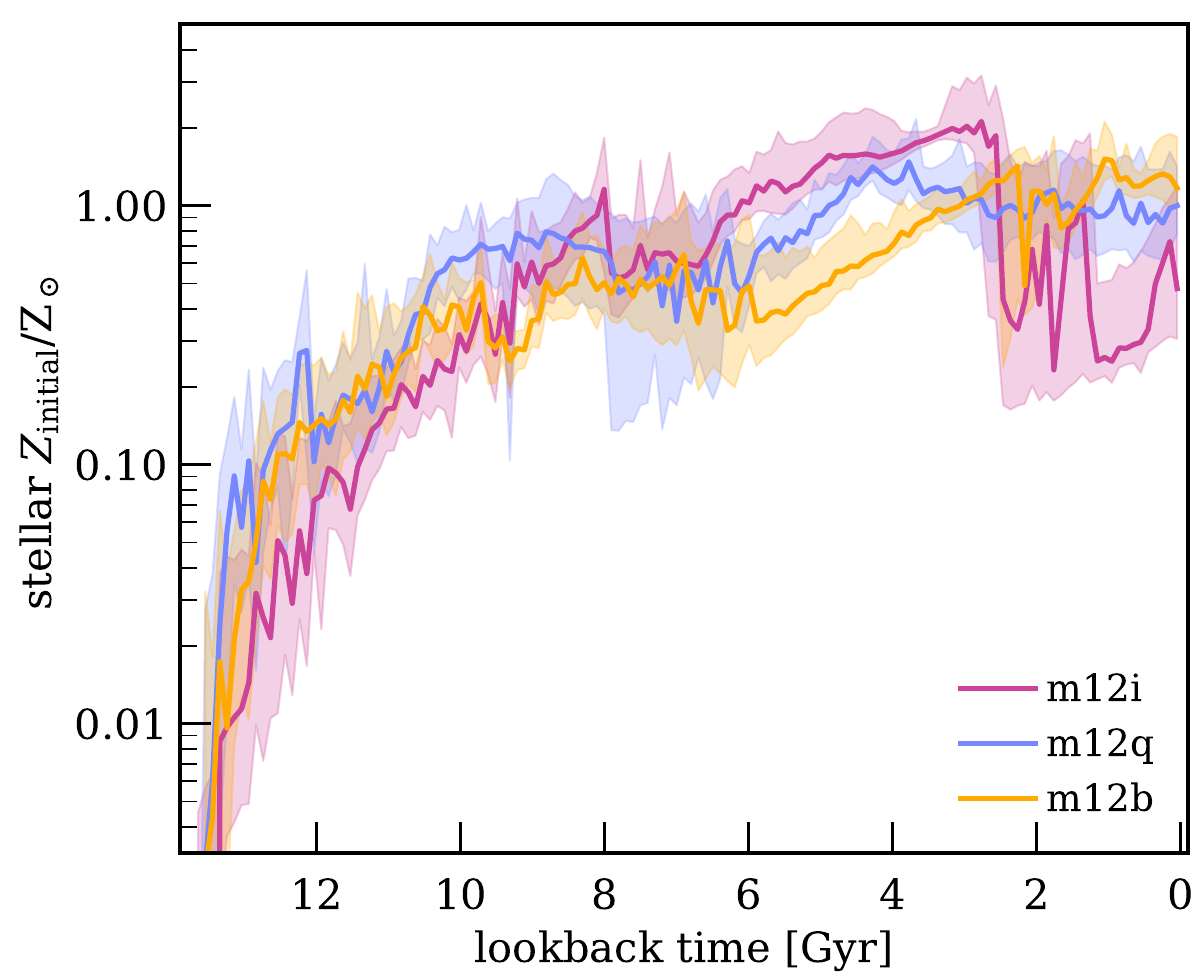} 
\end{tabular}
\caption{
Same as the left panel of Figure~\ref{fig:comparing-enrichment-histories}, but in log space and normalized to Solar metal mass fraction $Z_\odot$.
}
\label{fig:appendix-log-enrichment}
\end{figure}

Figure~\ref{fig:appendix-log-enrichment} is an alternative presentation of the left panel of Figure~\ref{fig:comparing-enrichment-histories}, in log space and normalized to Solar abundance $Z_\odot$, which might be a more intuitive presentation for some readers. Same takeaways as discussed in Section~\ref{sec:3.x-SFH-comparison}.

\bibliography{helium-metals}{}
\bibliographystyle{aasjournalv7}



\end{document}